\documentclass[12pt, draft, journal, onecolumn]{IEEEtran}
\usepackage[utf8]{inputenc}
\usepackage{amssymb,times,amsfonts,amsmath,amsthm,mathrsfs,
latexsym,mathtools}
\usepackage[final]{graphicx}

\usepackage{cite}

\newtheorem{theorem}{Theorem}
\newtheorem{proposition}{Proposition}
\newtheorem{lemma}{Lemma}
\newtheorem{corollary}{Corollary}
\newtheorem{example}{Example}
\newtheorem*{examples*}{Examples}
\newtheorem{remark}{Remark}

\newenvironment{proofoutline}{%
  \proof}{\endproof}

\newcommand{\expectation}{\ensuremath{\mathbb{E}}}
\newcommand{\Expt}{\expectation}
\newcommand{\probability}{\ensuremath{\mathbb{P}}}
\newcommand{\Prob}{\probability}
\def\fF{\mathbb{F}}

\def\cY{\mathcal{Y}}

\def\cY{\mathcal{Y}}

\def\<{\langle}
\def\>{\rangle}

\newcommand{\ssep}{\,\colon\,}

\DeclarePairedDelimiterX{\infdivx}[2]{(}{)}{%
  #1\delimsize\| #2%
}

\DeclareMathOperator{\arctanh}{arctanh}

\begin{document}

\title{Channel capacity of polar coding with a given polar mismatched successive cancellation decoder}

\author{
  \IEEEauthorblockN{Mine Alsan}\\
  \IEEEauthorblockA{  Department of Electrical and Computer Engineering\\
        			       National University of Singapore, Singapore\\ 
			       Email: minealsan@gmail.com
}}



\maketitle
\IEEEpeerreviewmaketitle

{\let\thefootnote\relax\footnotetext{The material in this paper was presented in part at the IEEE Information Theory Workshop, Sevilla, Spain, 2013, and at the IEEE Information Theory Workshop, Hobart, Australia, November 2014.}}

\begin{abstract}
Ar{\i}kan's polar coding is by now a well studied technique that allows
achieving the symmetric capacity of binary input memoryless channels with
low complexity encoding and decoding, provided that the polar decoding
architecture is used and the decoding metric is matched to the true channel. In this paper, we analyze communication rates that are achievable when the polar coding/decoding architecture is used with the decoder using an incorrect model of the channel. We define the ``polar mismatched capacity" as an analogue of the classical mismatched capacity, give an expression for it, and derive bounds on it. 
\end{abstract}


\section{Introduction}
In various communication scenarios, we encounter sub-optimal decoders due to partial or missing channel information or practical implementation constraints. To give an example of an obstacle on the way of optimal decoding, 
we can consider the case where a high signal to noise ratio channel is used with a large constellation with points indexed by $k$-bit symbols $s(0...0),...,s(1...1)$, and the receiver is interested in recovering the first of these $k$ bits.
Then, the true likelihood ratio requires the computation of the sums
\begin{equation*}
  \sum_{b_2,...,b_k} W(y|s(0,b_2,...,b_k)) \quad \hbox{ and} \quad
  \sum_{b_2,...,b_k} W(y|s(1,b_2,...,b_k)),
\end{equation*}
each containing an exponential number of terms (in $k$).
The receiver hardware may not permit such computations, and the decoder designer may be forced to use a simpler metric $V(y|1)/V(y|0)$ which approximates the true one.
In such cases, even when the receiver is informed of the true channel $W$, the decoding operation proceeds on the basis of a mismatched channel $V$. Regardless of the nature of the obstacle, sub-optimal decoders might perform worse than optimal decoders and result in capacity loss. Modeling such sub-optimal scenarios in the framework of \textit{`reliable communication with a given decision rule'} and establishing coding theorems for them allows one to assess the extent of any loss. 

Due to their many desirable properties, among which are good performance and low complexity, polar codes are being considered as candidate error correction codes for use in future wireless communication systems. The ideas behind polar coding, which we will discuss later in details, make successive cancellation decoding of polar codes the choice of decision rule for developing an elegant yet practical theory. Although this is a sub-optimal decision procedure, most of the results developed so far on polar coding,  including the initial theory in \cite{1669570} and multiple extensions to it, make the assumption that the scheme employs  successive cancellation decoders operating with metrics matched to the true channel laws. The analysis provided in \cite{1669570} shows that it is possible to achieve the symmetric capacity of binary input discrete memoryless channels (B-DMCs) using the polar coding scheme, and thus, the use of a successive cancellation decoder does not result in any capacity loss. 

On the other hand, as just discussed, it may not be possible for the implemented decoder to make its computations based on a decoding metric that is exactly matched to the true channel due to complexity or feasibility requirements. Such limitations bear important engineering implications for system designers: take for instance the work in \cite{380049} which investigates the problem of optimally assigning a mismatched metric to the received signal from a finite set of metrics, driven by the consideration that the receiver implementation will not permit infinite-precision arithmetic. Motivated by these observations, the primary objective of this paper is to study the effects of a decoding mismatch on the transmission capacity of polar coding over B-DMCs. 

\subsubsection*{Paper Outline}
In the rest of this section, we first introduce some basic notations used throughout the paper, and then we briefly survey the literature on the subject of reliable communication with a given decision rule.  Next, we provide the preliminaries about polar coding and we introduce the definition of the mismatched polar decoders in Section \ref{sec:pre-Polar-Coding}. After that, we introduce in Section \ref{sec:pre-Mismatched-Parameters} the preliminaries on the mismatched channel parameters that play an important role in the exposition of the results and their proofs.  Then, we describe the main results of this paper in Section \ref{sec:Overview} through Theorems \ref{thm:mismatched_polarization_D}--\ref{thm:rate_of_polarization_new}, and subsequently, we prove the theorems in Section \ref{sec:polar_mismatched_capacity}. Finally, we conclude the paper with some final remarks in Section \ref{sec:final_remarks}.

\subsection{Preliminary Notations}
Let $W:\mathcal{X}\to\mathcal{Y}$ denote a discrete memoryless channel (DMC) with input alphabet $\mathcal{X}$, output alphabet $\mathcal{Y}$, and transition probabilities $W(y|x)$, for $x\in\mathcal{X}$ and $y\in\mathcal{Y}$. When $W$ is a B-DMC, we will use $\mathbb{F}_2 := \{0, 1\}$ to denote the input alphabet of the channel. When the distribution of a random variable $X$ is $P(x)$, for $x\in\mathcal{X}$, this is denoted as $X\sim P(x)$. Given a random variable $Q$ defined on a probability space $(\Omega, \mathscr{F}, \Prob)$, we define
$\Prob(\{\omega\in\Omega: Q(\omega)=q\}) := \displaystyle\sum_{\omega\in\Omega} \Prob(\omega)\mathbf{1}\{Q(\omega)=q\}$. For a given B-DMC $W:\mathbb{F}_2\to\mathcal{Y}$,  we will compute probabilities of the form $W(\{y \in\mathcal{S}\}|u) := \displaystyle\sum_{y\in\mathcal{S}} W(y|u)$, for $u\in\mathbb{F}_2$ and $\mathcal{S}\subseteq\mathcal{Y}$. 
As customary, the expectation operator for random variables is denoted by $\Expt[.]$. We use the notation $|x|^+ = \max\{x,0\}$. The notation $\log$ stands for the base 2 logarithm, and $\ln$ stands for the natural logarithm. The set of non-negative integers is denoted by $\mathbb{N}$. Finally, standard Landau notations $O(\cdot)$, $o(\cdot)$, and $\omega(\cdot)$ are used to describe the asymptotic behavior of functions.

\subsection{Reliable communication with a given decision rule}\label{sec:mismatch_pre}
Shannon's treatment \cite{6773024} of the noisy channel coding theorem, or Gallager's refined error analysis \cite{578869}, both use decoders (typicality and ML decoders, respectively)  which require the exact channel knowledge to function. As a consequence, the scope of their results does not cover the case where this knowledge is missing. In contrast, the method of types of Csisz{\'a}r et. al.~\cite{CKM}, see also \cite[Thm. IV.1]{720546}, yields a universal coding theorem for DMCs by relying on the maximum mutual information (MMI) decoder, which was initially introduced by Goppa~\cite{MMI:Goppa}. Since the MMI decoder operates with empirical observations instead of true channel laws, the analysis applies to situations where this law is unknown to the code designer. Even though the MMI decoder is robust, it comes short in practice due to its highly complex computational nature. 

Considerations of decoding complexity has motivated the development of the ``mismatched decoding literature" devoted to the study of the transmission capacity of channels for the family of decoders with additive decision rules. Csisz\'{a}r and Narayan \cite{394641} studied the performance of this family of decoders called $d$-decoders. A $d$-decoder computes its decision function using the additive extension of a single letter metric $d(x, y)$, for $x\in\mathcal{X}$ and $y\in\mathcal{Y}$. The transmission capacity of a DMC $W:\mathcal{X}\to\mathcal{Y}$ when decoded with an additive metric $d$ is denoted by $C_d(W)$, see \cite[Def. 1]{394641}.  When the metric $d$ corresponds to the likelihood with respect to a mismatched channel $V$, the $d$-decoder is called a \textit{mismatched ML decoder} and $C_d(W)$ is called the \textit{mismatched capacity} of the channel $W$. In this paper, we denote the mismatched capacity by $C(W, V)$.  It is well-known that the ``matched capacity" of a DMC $W$, denoted by $C(W):= C(W, W)$, is given by the  single-letter expression $C(W) = \max_{P(x)}I(P,W)$, in which the maximization is over all possible channel input distributions and the mutual information between the channel input $X\sim P(x)$ and output $Y$ is given by
\begin{equation}
I(P,W)=\displaystyle\sum_{x\in\mathcal{X}}\displaystyle\sum_{y\in\mathcal{Y}} P(x)W(y|x)\log\left(\displaystyle\frac{W(y|x)}{\displaystyle\sum_{x'\in\mathcal{X}}P(x)W(y|x').}\right).
\end{equation}
On the other hand, no general closed form single letter expression is known for $C(W,V)$ or $C_d(W)$. Single-letter lower bounds have been derived, but no converse for any of the lower bounds exists, except for some very special cases. The class of binary-input binary-output channels is such a special case where either $C_{d}(W) = C(W)$ or $C_{d}(W) = 0$ hold depending on whether or not the mismatch metric is in `harmony' with the channel behavior \cite{394641}. Until recently, another exception was thought to be the class of B-DMCs. Balakirsky  \cite{Balakirsky}, \cite{476314} claimed a converse and gave a computable expression for $C_d(W)$ when $W$ is a B-DMC. However, a recent result of Scarlett et. al. \cite{2015arXiv150802374S} casts doubt on the veracity of this converse by providing a numerical counter-example.
  
Hui derived in \cite[Chap. 4]{Hui/THESES} a single-letter lower bound on $C(W, V)$. This lower bound turned out to coincide with the special case of a more general lower bound derived earlier in \cite{1056281} for the larger class of $\alpha$-decoders. This is a family of decoders accommodating decision rules based on the joint type of the codewords and the channel output sequences,  such as the MMI decoding rule. An in-depth study of the properties of this lower bound, carried out later in  \cite{340469}, introduced the notation $C_{\mathrm{LM}}$ to denote this lower bound. Dubbed as the ``LM rate'', $C_{\mathrm{LM}}$ became the generally accepted term in the ensuing literature to refer to the lower bound on $C(W, V)$ given in \cite{Hui/THESES, 1056281}. It is shown in \cite{340469} that if one insists on using a random coding argument, $C_{\mathrm{LM}}$ becomes the highest possible achievable rate for mismatched ML decoding, i.e., the channel capacity of random coding with a given mismatched ML decoder. Improvements over the LM rate have been also studied in the literature. The simplest method is to recourse to the product space and to apply the arguments that led to the derivation of $C_{\mathrm{LM}}$ to the super-channel corresponding to the $N$ independent uses of the original DMC. It has been conjectured that as $N\to\infty$, the family of lower bounds tend to $C(W, V)$. However, it is in general not possible to compute $C_{\mathrm{LM}}$ corresponding to a super-channel when $N$ is large. Another improvement over the lower bound $C_{\mathrm{LM}}$ was also obtained by Lapidoth \cite{532884} which derived a single letter expression by treating the single-user channel as a multiple-access channel. The earlier literature on mismatched decoding is broader and includes other works such as \cite{1054129, Pinsker-Sheverdyaev, 394640, 481778, 532892, 887846}. For instance, \textit{erasures only metrics} are special cases of $d$-decoding where $d(x, y) = 0$ if and only if $W(y|x)>0$. For more recent developments, see \cite{7134788}, \cite{Scarlett/THESES}, and the references therein.  

The problem setup we are interested in this paper is different and new. In the classical mismatched capacity problem, one fixes the true channel $W$ and a mismatched channel $V$ used by the ML decoder, and then designs a code with the full knowledge of $W$ and $V$. The classical mismatched capacity is the highest rate for which a reliable code may be designed. In this paper, we take an analogous viewpoint, assume full knowledge of $W$ and $V$, but insist on using polar codes and polar decoders as opposed to arbitrary codes and ML decoders, and we study the highest rate for which such a reliable code may be designed. We call this rate the \textit{polar mismatched capacity}. Note that polar decoders can be considered as another decoder family based on successive cancellation decoding procedures. They offer a quite different decoding paradigm than additive decoders. A successive cancellation decoder decodes the received output in multiple stages using a chain of estimators, each possibly depending on the previous ones. The successive cancellation decoder used by the polar coding scheme is particularly attractive for yielding coding theorems proving the symmetric capacity achieving property of the scheme with a low complexity architecture \cite[Thm. 5]{1669570}.

\section{Preliminaries on Polar Coding}\label{sec:pre-Polar-Coding}


In this section, we review the original polar coding method proposed in \cite{1669570} for B-DMCs as this method will play a crucial role in the exposition and understanding of the results in this paper. In addition, we review the dominant analysis technique used to prove coding theorems in the context of polar coding. Lastly, we present the problem setup we study by introducing the definition of the mismatched polar successive cancellation decoder. 

\subsection{Channel Polarization and Polar Codes}
In his seminal paper \cite{1669570}, Ar{\i}kan introduced polar coding as a method to construct a family of error correcting codes achieving the symmetric capacity of B-DMCs by taking advantage of a phenomenon called \textit{channel polarization}. The original and subsequent works showed that this family of codes, named as polar codes, possess many desirable properties that make them amenable to practice and set them apart from the state-of-the art coding techniques such as LDPC and Turbo codes. As we now review, polar codes are defined with an explicit construction, require low complexity encoders and decoders, and have analytical error probability bounds.

Suppose that we have access to multiple copies of a given B-DMC $W: \mathbb{F}_2\to \mathcal{Y}$ whose symmetric capacity, i.e., mutual information evaluated for the uniform input distribution, is denoted by $I(W)$. Consider the super-channel $W^N:  \mathbb{F}_2^N \to \mathcal{Y}^N$ formed by the use of $N$ copies of this channel with independent and uniformly distributed binary inputs. The channel polarization effect is observed in  \cite{1669570} by synthesizing from $W^N$ a set of $N$ channels that exhibit a certain asymptotic behavior: as $N$ is increased, the synthesized channels polarize into two clusters, having in one group almost perfect channels with symmetric capacities close to 1, and in the other, almost completely noisy channels with symmetric capacities close to 0, for all but a vanishing fraction of channels. The channels synthesized from $W^N$ are denoted by $W_N^{(i)}: \mathbb{F}_2\to \mathcal{Y}^N\times  \mathbb{F}_2^{(i-1)}$, for $i=1, \ldots, N$. The transition probabilities of the channel $W_N^{(i)}$ and the super-channel $W^N$ are related by the following expression \cite[Eq. (5)]{1669570}: 
\begin{equation}\label{eq:synthetic-channel-prob}
W_{N}^{(i)}(y_{1}^{N} u_{1}^{i-1}\mid u_{i}) =  \displaystyle\sum_{u_{i+1}^N \in \mathbb{F}_2^{N-i}} \frac{1}{2^{N-1}} W^N(y_1^N|u_1^NB_{N} F^{\otimes n}),
\end{equation}
where $N=2^n$, for $n\in\mathbb{N}$,  $B_{N}$ is a permutation matrix known as bit-reversal,  the recursion $F^{\otimes n}$ is the $n$-th recursive Kronecker product of the matrix
\begin{equation} \label{eq:F}
F := \left[ \begin{array}{cc}
			1 & 0 \\ 1 & 1
           \end{array}\right],
\end{equation}
and the transition probabilities of the super-channel is given by $W^N(y_1^N|x_1^N) = \prod_{j=1}^{N}W(y_{j}|x_{j})$. The channel polarization phenomenon is demonstrated in \cite{1669570} by proving the following theorem.
\begin{theorem} \cite[Thm 1]{1669570} \label{thm:polarization}
Let $N=2^n$ with $n\in\mathbb{N}$. For any B-DMC $W$ and for all $ \gamma\in(0, 1)$, 
  \begin{equation} \label{eq:polarization-1} 
\frac{1}{N} \hspace{2mm} \# \{i\in\{1, \ldots, N\}\ssep I(W_{N}^{(i)}) \in (\gamma, 1- \gamma)\}  \xrightarrow[n\to\infty]{}  0, 
\end{equation}
such that
\begin{equation} \label{eq:polarization-2} 
\frac{1}{N} \hspace{2mm} \# \{i\in\{1, \ldots, N\}\ssep I(W_{N}^{(i)}) \in (1-\gamma, 1]\} \xrightarrow[n\to\infty]{} I(W), 
\end{equation}
\begin{equation} \label{eq:polarization-3} 
\frac{1}{N} \hspace{2mm} \# \{i\in\{1, \ldots, N\}\ssep I(W_{N}^{(i)}) \in [0, \gamma)\} \xrightarrow[n\to\infty]{} 1-I(W).
\end{equation}
\end{theorem}

Let us now explore the implications of the transformation in \eqref{eq:synthetic-channel-prob} from a channel coding perspective. This transformation can be viewed as a two step procedure. The first step consists of a channel combining operation which generates the inputs to the channel $W^N$ via the generator matrix $G_{N} :=  B_{N} F^{\otimes N}$. Applied as $x_1^N =u_1^NG_N$, for a given message $u_1^N\in\mathbb{F}^N$, the generator matrix results in an intermediary channel $W_N: \mathbb{F}_2^N \to \mathcal{Y}^N$ whose transition probabilities are given by $W_N(y_1^N|u_1^N) = W^N(y_1^N|u_1^NG_N)$ \cite[Eq. (4)]{1669570}.  The second step consists of a channel splitting operation, which re-splits the combined channels, as described in \eqref{eq:synthetic-channel-prob}, into a final set of $N$ ``successive'' channels  $W_N^{(i)}$ with input $u_i$  and outputs $y_1^N u_1^{(i-1)}$, for $i=1, \ldots, N$. 

We pause here for a moment to introduce important terminology. Having the channel $W_N$ at hand, one can, in principle, encapsulate a sequence of $K$-bits of information within the input $u_1^N$ of the channel $W_N$, assuming $K\leq N$, by placing each bit of information into some $K$ coordinates of  $u_1^N$, and by fixing the values of the remaining $(N-K)$ coordinates to some arbitrary known values. The choice of the $K$ coordinates will determine which $K$ rows of the matrix $G_N$ will be used for encoding the $K$-bits of information before transmission over $W^N$. Coding in this context is described by the following terminology: the set of indices corresponding to the selected  $K$ rows of $G_N$ is called the information set and is denoted by $\mathcal{A}_N \subseteq \{1, \dots N\}$, the remaining set of indices is called the frozen set and is denoted by $\mathcal{F}:=  \{1, \dots N\}\setminus\mathcal{A}_N$, and the vector of fixed values $u_{ \mathcal{F}}$ are called the frozen inputs. In the coding theory literature, an error correcting code which encodes  $K$-bits of information into an $N$-length input to  $W^N$ in this way  is called an $(N, K, \mathcal{A}_N,  u_{ \mathcal{F}})$-$G_N$-coset code of rate $K/N$. We shall emphasize that in this context, the information set and the frozen inputs are known at both the transmitter and receiver sides. 

We are now ready to describe how the polar coding scheme works. The two channel operations result in a communication setting (for large blocklengths) where one has access to the binary inputs $u_1^N$ of a set of synthetic channels polarized in the sense of Theorem \ref{thm:polarization}. Polar coding defines a polar code as an-$(N, K, \mathcal{A}_{N, \gamma}(W),  u_{\mathcal{F}})$-$G_N$-coset code with the following choice of the information set:
\begin{equation}\label{eq:info_set_original}
\mathcal{A}_{N, \gamma}(W) := \{i\in\{1, \ldots, N\}\ssep I(W_{N}^{(i)}) \geq 1-\gamma \},
\end{equation}
where $\gamma\in(0, 1)$ is a desired threshold. The polar encoder proceeds by computing the inputs $x_1^N$ as described in the previous paragraph, and transmits them over the super-channel $W^N$. Typically, uncoded transmission is optimal over an almost perfect channel since no information loss occurs during transmission. However, due to the channel splitting operation, the output of the channel $W_{N}^{(i)}$ depends both on the received channel outputs $y_1^N\in\mathcal{Y}^N$ and the previous inputs $u_{1}^{i-1}\in\mathbb{F}_2^{i-1}$. For this reason, polar codes lend themselves to a successive cancellation decoding procedure.  The polar successive cancellation decoder, or simply the polar decoder, decodes the received channel output using a chain of estimators $\hat{u}_{i}\in\mathbb{F}_2$, for $i=1, \ldots, N$, where  $\hat{u}_{i}$ is set to its known frozen value if $i\in\mathcal{F}$ and decoded otherwise by employing the ML decision rule of the corresponding synthetic channel \cite[Eq.s (11) \& (12)]{1669570}.  For a given B-DMC $W$, let $P_{\textnormal{e}}(W, \mathcal{A}_N)$ denote the best achievable block decoding error probability under polar successive cancellation decoding for a code optimized among the ensemble of $(N, K, \mathcal{A}_{N},  u_{ \mathcal{F}})$-$G_N$ codes over all choices of the frozen input values for fixed $\mathcal{A}_{N}$. We know by \cite[Prop. 19]{1669570}, \cite[Thm. 2]{5205856}  that polar coding performs well:  For any fixed rate $R< I(W)$, and any constant $\beta\in(0, 1/2)$,  any information subset $\mathcal{A}_{N}^* \subseteq\mathcal{A}_{N, \gamma}(W)$ of size $\lfloor RN\rfloor$ selected according to the definition in \eqref{eq:info_set_original} 
 satisfies ${P_\textnormal{e}}(W, \mathcal{A}_{N}^*)  = o(2^{-N^\beta})$ for the threshold $\gamma=o(2^{-N^\beta})$.
 
\subsection{A note on the dominant analysis technique used to prove polar coding theorems}\label{sec:pre-analysis}

Here, we present some of the key aspects pertaining to the analysis technique used in the original proofs of Theorem \ref{thm:polarization} and  the coding theorem stated in \cite[Prop. 19]{1669570}, \cite[Thm. 2]{5205856} as we will be using a similar analysis. One of the key ingredients in proving these results is to exploit the recursive structure inherent to the generator matrix $G_N$. It is customary to refer to the one-step application of the basic matrix $F$, defined in \eqref{eq:F}, as the polar transform. For analysis purposes, it is also convenient to index the synthetic channels via sequences of plus and minus signs. Next, we define the polar transform and the synthetic channels using this alternative labeling.  Throughout the paper, we alternate between the two notations for convenience. 

From two independent copies of a given B-DMC $W$, the polar transform synthesizes two new B-DMCs denoted as $W^-:= W_2^{(1)} \colon\fF_2\to\cY^2$ and $W^+:= W_2^{(2)} \colon\fF_2\to\cY^2\times\fF_2$. The transition probabilities of these channels are given by \cite[Eqs. (19) \& (20)]{1669570} 
\begin{align}
\label{eq:def_pol_minus}&W^-(y_1y_2|u_1) :=\sum_{u_2\in\fF_2}\displaystyle\frac{1}{2} W(y_1|u_1\oplus u_2)W(y_2|u_2),\\
\label{eq:def_pol_plus}&W^+(y_1y_2u_1|u_2) :=\displaystyle\frac{1}{2} W(y_1|u_1\oplus u_2)W(y_2|u_2).
\end{align}
Some basic properties of the mapping $(W, W) \to (W^-, W^+)$ are established in \cite[Prop. 4]{1669570}. Key properties of the recursive application of the polar transform can be inferred from its one-step properties by introducing an auxiliary stochastic process called the channel polarization process\footnote{For an elementary proof of Theorem \ref{thm:polarization} which does not introduce an auxiliary random process, see  \cite[Proof of Thm. 1]{7496994}.}.

Let $\Omega:=\{0, 1\}^{\infty}$ be the space of infinite binary sequences $w:= (w_1, w_2, \ldots)$. For each $n\in\mathbb{N}$, let $\mathscr{F}_n$ be the $\sigma$-field generated by the cylinder sets $\{w\in\Omega: w_1=b_1, \ldots, w_k=b_k\}$, for $k\in\{1, \ldots, n\}$ and $b_1, \ldots, b_k\in\{0, 1\}$. Consider the probability space $(\Omega, \mathscr{F}_{\infty}, \Prob)$, where $\Prob\left[\{w\in\Omega: w_1=b_1, \ldots, w_n=b_n\}\right]:= 1/2^n$ holds for all $n\in\mathbb{N}$ and $b_1, \ldots, b_n\in\{0, 1\}$. We first define on this space the process $B_n(w) := w_n$, for $w\in\Omega$ and $n\in\mathbb{N}$. Notice that $\{B_n\}_{n\in\mathbb{N}}$ is a sequence of i.i.d. Bernoulli random variables drawn according to the Bernoulli distribution with probabilities equal to 1/2, and each set of the form $\{w\in\Omega: w_1=b_1, \ldots, w_n=b_n\}$ corresponds to the event that the collection $B_1, \ldots, B_n$ takes the specific value $b_1, \ldots, b_n\in\{0, 1\}$. Following this observation, the channel polarization process is defined by letting \cite{1669570}, \cite{arXiv:Rate-of-pol} 
\begin{equation}\label{eq:def_pol_process}
W_{n+1}:=\begin{cases}W_n^-,& \hbox{if } B_{n+1} = 0\\
W_n^+,&\hbox{if } B_{n+1} = 1\end{cases}, 
\end{equation}
for $n\geq 1$, and $W_0:=W$. In this way, the repeated application of the polar transform yields, at stage $n\geq 1$, a channel random variable $W_n$ which is uniformly distributed over the set of $2^n$ synthetic channels $\bigl\{W^{s^n} \ssep s^n\in\{+,-\}^n\bigr\}$. For a given realization $b_1, \ldots, b_n$ of the Bernoulli collection $B_1, \ldots, B_n$, while the index of the synthetic channel $W_N^{(i)}$ is given by $i :=  1 + \sum_{j=1}^{n}b_j2^{n-j}$, see \cite[Sec. IV]{1669570}, the sign sequence $s^n:= (s_1, \ldots, s_n)$ identifying this channel in the notation $W^{s^n}$ is determined by mapping $s_j = -$, if $b_j=0$, and $s_j=+$, otherwise, 
 for $j\in\{1, \ldots n\}$. Once the channel polarization process is defined, one can further introduce other auxiliary random processes that follow the evolution of information measures as the underlying channel undergoes the sequence of polar transforms. Two such measures are given by the symmetric capacity process $I_n(W)  := I(W_n)$ and the Bhattacharyya parameter process $Z_n(W)  := Z(W_n)$, where the Bhattacharyya parameter of a B-DMC $W$ is given by $Z(W)=\sum_{y\in\mathcal{Y}} \sqrt{W(y|0)W(y|1)}$. These processes are $\mathscr{F}_{n}$-measurable for any fixed $n\in\mathbb{N}$. 
 
The usefulness of this formulation is demonstrated as follows. It is shown in\cite[Props. 8 \& 10]{1669570} that the process $I_{n}(W)$ is a bounded martingale on the interval $[0, 1]$, and furthermore, it converges almost surely (a.s.) towards a limiting random variable $I_{\infty}:= \lim_{n\to\infty} I_n(W)$ such that $I_{\infty}\in\{0, 1\}$ and $\Expt{\left[ I_{\infty}\right] } = I_{0}$. Thus, Theorem \ref{thm:polarization} is proved by appealing to the theory of martingales. To prove the coding theorem stated in \cite[Prop. 19]{1669570}, \cite[Thm. 2]{5205856}, it is first observed that \cite[Prop. 2]{1669570}
\begin{equation}\label{eq:Pe-upper-bound}
P_{\textnormal{e}}(W, \mathcal{A}_N)   \leq  \displaystyle\sum_{i\in\mathcal{A}_{N}} Z(W_{N}^{(i)}) \leq N\displaystyle\max_{i\in\mathcal{A}_{N}} Z(W_{N}^{(i)})
\end{equation}
holds. Since by \cite[Props. 1 \& 11]{1669570}, the limiting random variable $Z_\infty:= \lim_{n\to\infty}Z_n$ exists a.s., and $Z_{\infty}=0$ holds if and only if $I_{\infty}=1$, and vice versa, the proof of the coding theorem is then completed by showing that ``whenever the process $Z_n(W)$ converges to zero, this convergence is a.s. fast''.\footnote{From \eqref{eq:Pe-upper-bound},  we can see that the ``fast polarization'' phenomenon alludes to the convergence of a process associated to $W_n$ toward the value of the parameter attained by an almost perfect channel at a rate faster than $1/N$.} At last, we re-state this ``fast polarization theorem'' proved in \cite{5205856}\footnote{Initially, Theorem \ref{thm:rate_of_polarization_Qn} and its proof appeared in \cite{arXiv:Rate-of-pol}. The proof in \cite{5205856}, which is of interest to us in this paper, was published later to present a simplified version of the proof given in \cite{arXiv:Rate-of-pol}. More recently, the claim of the theorem has been slightly generalized and its proof further simplified in \cite[Lemma 2]{8052539}.} in the context of a more general branching process which includes the process  $Z_n(W)$ as a special case\footnote{ Note that this branching process need not be associated to the channel polarization process of any B-DMC.} . 

\begin{theorem} \cite[Thms. 1 \& 3]{5205856}\label{thm:rate_of_polarization_Qn}
Let $Q_ n$ be a process such that:
\begin{itemize}
\item[$(c.1)$] For each $n \in\mathbb{N}$, $Q_n\in[0, 1]$, $Q_0$ is
constant, and $Q_n$ is measurable with respect to $\mathscr{F}_n$.
\item[$(c.2)$] For some constant $q\geq 2$ and for each $n\in \mathbb{N}$,
\begin{align}
&Q_{n+1} = Q_ n^2,  &\quad \hbox{when } B_{n+1} = 1,\\
&Q_{n+1} \leq qQ_n, &\quad \hbox{when } B_{n+1} = 0.
\end{align}
\item[$(c.3)$] $Q_\infty:= \lim_{n\to\infty} Q_n$ exists a.s. and $Q_\infty$ is a $\{0, 1\}$-valued random variable. 
\end{itemize}
Then, for any $\beta\in(0, 1/2)$,
\begin{equation}\label{eq:Qn_convergence_1}
\lim_{n\to\infty}\Prob[ Q_n < 2^{-2^{n\beta}}] = c,
\end{equation}
where $c:=\Prob(Q_\infty = 0)\in[0, 1]$.
Moreover, if the condition $(c.2)$ is replaced by the condition
\begin{itemize}
\item[$(\tilde{c}.2)$] For each $n\in \mathbb{N}$,
\begin{align}
&Q_{n+1} = Q_ n^2, &\quad \hbox{when } B_{n+1} = 1,\\
&Q_{n+1} \geq Q_n, &\quad \hbox{when } B_{n+1} = 0,
\end{align}
\end{itemize}
then, if $Q_0 > 0$, for any $\beta > 1/2$,
\begin{equation}\label{eq:Qn_convergence_2}
\lim_{n\to\infty}\Prob[ Q_n < 2^{-2^{n\beta}}] = 0.
\end{equation}
\end{theorem}
We will essentially use Theorem \ref{thm:rate_of_polarization_Qn} in proving our polar coding theorem with mismatched polar decoding. 

\subsection{Mismatched Polar Successive Cancellation Decoder}
The mismatched polar successive cancellation decoder estimates the messages encoded by an $(N, K, \mathcal{A}_N,  u_{ \mathcal{F}})$-$G_N$-coset code by running a multi-stage decision procedure. Upon reception of a sequence $y_1^N$ at the output of the channel $W^N$,  the decoder successively declares $\hat{u}_{i} = u_{i}$, if $i\in\mathcal{F}$, and $\hat{u}_{i} = d_{V_N^{(i)}}(y_{1}^{N}, \hat{u}_{1}^{i-1})$, if $i\in\mathcal{A}_{N}$, in the order from 1 to $N$, by employing the following decision functions 
\begin{equation}\label{eq:polar_SCD_decision_rule}
d_{V_N^{(i)}}(y_{1}^{N}, \hat{u}_{1}^{i-1}) := \left\{\begin{array}{ll}
							0, & \hbox{if} \hspace{1mm} L_{V_{N}^{(i)}}(y_{1}^{N}, \hat{u}_{1}^{i-1}) < 1 \\
							1, & \hbox{if} \hspace{1mm} L_{V_{N}^{(i)}}(y_{1}^{N}, \hat{u}_{1}^{i-1}) > 1\\
							*, & \hbox{if} \hspace{1mm} L_{V_{N}^{(i)}}(y_{1}^{N}, \hat{u}_{1}^{i-1}) = 1
						       \end{array}  \right.,
\end{equation}
where $*$ is chosen from the set $\{0, 1\}$ by a fair coin flip, and the likelihood ratios are defined as $L_{V_{N}^{(i)}}(\cdot) = V_{N}^{(i)}(\cdot|1)\big/ V_{N}^{(i)}(\cdot|0)$.  Note that the mismatched polar decoder is structurally identical to the matched polar decoder introduced in \cite{1669570}. The difference is operational since the decision functions $d_{V_N^{(i)}}(y_{1}^{N}, \hat{u}_{1}^{i-1})$ do still apply the ML decoding rule but with respect to
mismatched channels $V_N^{(i)}$ synthesized by polarizing a B-DMC $V$ possibly different than the true communication channel $W$. Consequently, the $O(N\log{N})$ time complexity result given in \cite[Thm 5]{1669570} for the matched polar decoder is still valid. Furthermore, the mismatched polar decoder can be run efficiently by existing software/hardware implementations proposed in the literature for the matched polar decoder  \cite{1669570, 5946819, 7419248}.

\section{ Preliminaries on the Mismatched Channel Parameters}\label{sec:pre-Mismatched-Parameters}
In this section, we give the definitions of the mismatched channel parameters that will be encountered in the statements of the main results of this paper and their proofs. We also make few basic observations about these parameters which will be useful in later analyses.

For a given pair of B-DMCs $W$ and $V$, let $P_{\textnormal{e}}(W, V, \mathcal{A}_N)$ denote the best achievable block decoding error probability of a code from the ensemble of $(N, K, \mathcal{A}_N,  u_{ \mathcal{F}})$-$G_N$-coset codes under mismatched polar successive cancellation decoding with respect to the channel $V$ when the true channel is $W$. As in the mathced case, the code is optimized over all choices of the frozen input values for fixed $\mathcal{A}_{N}$.\footnote{Note that the definition of $P_{\textnormal{e}}(W, V, \mathcal{A}_N)$ allows sets $\mathcal{A}_{N}\subseteq\{1, \ldots, N\}$ which are not necessarily constructed as in \eqref{eq:info_set_original}.} 
We define
\begin{align}\label{eq:mismatch_ML_Pe}
P_{\textnormal{e, ML}}(W, V) := \displaystyle\frac{1}{2}\displaystyle\sum_{\begin{subarray}{c}  y\colon\\L_{V}(y)>1 \end{subarray}} W(y|0) 
&+ \displaystyle\frac{1}{4}\displaystyle\sum_{\begin{subarray}{c} y\colon\\L_{V}(y)=1 \end{subarray}}  W(y|0) \nonumber \\
&+  \displaystyle\frac{1}{2}\displaystyle\sum_{\begin{subarray}{c}  y\colon\\L_{V}(y)<1 \end{subarray}} W(y|1) 
+ \displaystyle\frac{1}{4}\displaystyle\sum_{\begin{subarray}{c}  y\colon\\L_{V}(y)=1 \end{subarray}}  W(y|1).
\end{align}
This quantity corresponds to the mismatched ML decoding error probability resulting from the single use of a B-DMC $W$ with uniformly distributed inputs when ML decoding with respect to a possibly mismatched B-DMC $V$ is used as a decoding metric. Similar to the analysis carried for the matched counterpart in \cite[Sec. V]{1669570}, we can upper bound $P_{\textnormal{e}}(W, V, \mathcal{A}_N)$  by:
\begin{align} {P_\textnormal{e}}(W, V, \mathcal{A}_{N})
&= \Prob\left[ \displaystyle\bigcup_{i\in\mathcal{A}_{N}} \left\lbrace \hat{U}_{1}^{i-1} = u_{1}^{i-1}, \hat{U}_{i} \neq u_{i} \right\rbrace\right] \nonumber\\
&= \Prob\left[\displaystyle\bigcup_{i\in\mathcal{A}_{N}} \left\lbrace\ \hat{U}_{1}^{i-1} = u_{1}^{i-1}, d_{V_N^{(i)}}(y_{1}^{N}, \hat{U}_{1}^{i-1}) \neq u_{i} \right\rbrace\right] \nonumber\\
&\leq \Prob\left[\displaystyle\bigcup_{i\in\mathcal{A}_{N}} \left\lbrace d_{V_N^{(i)}}(y_{1}^{N}, u_{1}^{i-1}) \neq u_{i} \right\rbrace\right] \nonumber\\
\label{eq:MSCD_Pe_UB} &\leq \displaystyle\sum_{i\in\mathcal{A}_{N}} {P_{\textnormal{e, ML}}}(W_{N}^{(i)}, V_{N}^{(i)}).
\end{align}
The parameters ${P_{\textnormal{e, ML}}}(W_{N}^{(i)}, V_{N}^{(i)})$ can be interpreted as the error probability of a ``genie-aided" mismatched maximum likelihood 
decoder for the $i$-th synthetic channel, where the genie provides the correct values of the previous inputs during each stage of the decoding procedure. Though no genie exists to give the correct estimates, thanks to this bound,  the analysis is simplified. 

Observe that, when $V\neq W$, even if $W_N^{(i)}$ and $V_N^{(i)}$ are both almost perfect channels having symmetric capacities close to 1, the mismatched ML decoding error probability ${P_{\textnormal{e, ML}}}(W_{N}^{(i)}, V_{N}^{(i)})$ itself may not be small; take for example, two binary symmetric channels (BSC) with crossover probabilities $\epsilon$ and $1-\epsilon$, for $\epsilon > 0$ small. Although both channels have high symmetric capacities, the error probability is also high. Consequently, when there is mismatch in decoding, it is not clear what data rates may be achieved by the polar coding/decoding architecture, and how to design polar codes in this setting. In an earlier study \cite{6401042}, the effects of introducing an approximation into the decoding procedure of the polar decoder has been considered for the class of BSCs.  The study also features simulation results on the error performance degradation introduced by a mismatched polar successive cancellation decoder at various blocklengths:  \cite[Fig. 3]{6401042} suggests that the error probability is still vanishing with $N\to\infty$ for mismatched pairs of BSCs. 

Ar{\i}kan \cite{1669570} chose to study the properties of the mapping $(W, W) \to (W^-, W^+)$ and its recursive application in terms of the symmetric capacity and the Bhattacharyya parameter for their interpretations as measures of rate and reliability, respectively, and also for their suitability for analysis purposes. We now proceed with the definition of the mismatched channel parameters which will be used in this paper to study the properties of the mapping $\left((W, V), (W, V)\right) \to \left((W^-, V^-), (W^+, V^+)\right)$. Given a B-DMC $W:\mathbb{F}_2\to\mathcal{Y}$, let us set
\begin{equation}\label{eq:dist}
q_W(y) = \frac{W(y\mid0) + W(y\mid1)}{2},
\end{equation}
and
\begin{equation}\label{eq:delta}
\Delta_W(y) = \frac{W(y\mid0)-W(y\mid1)}{W(y\mid0)+W(y\mid1)},
\end{equation}
for $y\in\mathcal{Y}$. Note that $q_W(y)$ is the output distribution of the channel when the inputs are uniformly distributed, and we have $W(y|0) = q_W(y)\left(1 + \Delta_{W}(y)\right)$ and $W(y|1) = q_W(y)\left(1 - \Delta_{W}(y)\right)$.\footnote{The polarization phenomenon for B-DMCs has been studied via the channel parameter $\Delta_W(y)$ in multiple chapters of \cite{Mine/THESES}.} Given two B-DMCs $W:\mathbb{F}_2\to\mathcal{Y}$ and $V:\mathbb{F}_2\to\mathcal{Y}$, the following two mismatched channel parameters are of primary interest to us in this paper: For $Y\sim q_W(y)$ with $y\in\mathcal{Y}$, we define
\begin{equation}\label{eq:mismatch_D_def}
D(W, V) := \Expt[\sqrt{\lvert\Delta_V(Y)\rvert}]  = \displaystyle\sum_{y\in\cY}q_W(y) \sqrt{\lvert\Delta_V(y)\rvert},
\end{equation}
and 
\begin{align}
\label{eq:mismatch_I_def} I(W, V) &:= \displaystyle\sum_{x\in\fF_2}\sum_{y\in\cY} \frac{1}{2} W(y|x) \log{\frac{V(y|x)}{q_V(y)}} \\
\label{eq:mismatch_I_def_delta} &= \frac{1}{2}\displaystyle\sum_{y\in\cY} W(y|0) \log\left(1 + \Delta_{V}(y)\right) + \frac{1}{2}\displaystyle\sum_{y\in\cY}W(y|1) \log\left(1 - \Delta_{V}(y)\right).
\end{align}
The other mismatched channel parameters that will be of auxiliary importance are given by 
\begin{equation}\label{eq:mismatched_var_distance}
 T(W, V) := \Expt[\lvert\Delta_V(Y)\rvert] = \sum_{y\in\cY} q_W(y) \lvert \Delta_V(y)\rvert,
 \end{equation}
   \begin{equation}\label{eq:mismatched_var_distance}
 T_{k}(W, V) := \Expt[\lvert\Delta_V(Y)\rvert^k] = \sum_{y\in\cY} q_W(y) \lvert \Delta_V(y)\rvert^k,
 \end{equation}
for any $k\in\mathbb{N}\setminus\{0, 1\}$, and 
\begin{equation}\label{eq:mismatched_Bhattacharyya}
Z(W, V) := \frac{1}{2}\displaystyle\sum_{y\in\cY}W(y|0) \sqrt{L_V(y)} + \frac{1}{2}\displaystyle\sum_{y\in\cY}W(y|1) \sqrt{\frac{1}{L_V(y)}},
\end{equation}
where we recall that $L_V(y)=V(y|1)/V(y|0)$ denotes the likelihood ratio of the channel $V$. For reasons that will become clear in the following three remarks, we call the parameters $I(W, V)$, $T(W, V)$, and $Z(W, V)$ as  the \textit{mismatched mutual information}, the \textit{mismatched variational distance}, and the \textit{mismatched Bhattacharyya parameter}, respectively. 

Even though the mismatched parameters might seem as pulled out of a magic hat, our choices of the mismatched channel parameters will be justified in Section \ref{sec:polar_mismatched_capacity} by the properties of their local transformation under the polar transform. All in all, the identification of the specific mismatched channel parameters suitable for the performance analysis of the mismatched polar decoder shall be acknowledged as part of the contributions of this paper. In the next three subsections, we explore the connections of the introduced mismatched channel parameters with the mismatched decoding literature, and we observe some of their basic properties.

\subsection{On the mismatched mutual information}\label{subsec:pre-MMI}
 Let us first start by mentioning that the definition of $I(W, V)$ equals the \textit{generalized mutual information} definition under the uniform input distribution in Fischer~\cite{Fischer}.\footnote{Kaplan and Shamai~\cite{kaplan1993information} derived a more general version of Fischer's generalized mutual information by using Gallager's error exponent derivation technique \cite{578869}, where one introduces a positive variable over which one should optimize. See also Merhav et. al.~\cite{340469} for the definition of this more general expression.} It is shown in \cite{Fischer} that the generalized mutual information is a lower bound on the mismatched capacity, i.e.,  $C(W, V) \geq I(W, V)$ holds, see also \cite[Rem. $i)$]{394641}. As $I(W, V)$ can take on negative values, $\lvert I(W, V)\rvert^+ = \max\{I(W, V), 0\}$ is also a lower bound on the mismatched capacity and an achievable rate with mismatched ML decoding. 
 
 In addition, we now show that $I(W, V)$ can be used to derive an upper bound on the mismatched ML decoding error probability in \eqref{eq:mismatch_ML_Pe}.
 \begin{proposition}\label{prop:rem1-p2}
 Consider the single use of a B-DMC $W\colon \fF_2\to\mathcal{Y}$ to transmit a $0$ or $1$. When both inputs are equally likely, the average mismatched ML decoding error probability  can be upper bounded as
\begin{equation}\label{eq:Pe_I_mismatch_UB}
P_{\textnormal{e, ML}}(W, V)  \leq \min\{1-I(W,V), 1\},
\end{equation}
for any B-DMC $V\colon \fF_2\to\mathcal{Y}$.
 \end{proposition}
 \begin{IEEEproof}
Given that the symbol $0$ is transmitted, one can derive the following upper bound to the error probability resulting from such a transmission:
\begin{align}\label{eq:Pe-ML-bound-0}
&P_{\textnormal{e, ML}}(W, V|0) := \displaystyle\sum_{\begin{subarray}{c}
y\colon\\L_{V}(y)>1 \end{subarray}} W(y|0) + \displaystyle\frac{1}{2}\displaystyle\sum_{\begin{subarray}{c}
y\colon\\L_{V}(y)=1 \end{subarray}}  W(y|0)  \nonumber\\
\leq &\displaystyle\sum_{\begin{subarray}{c}
y\colon\\L_{V}(y)\geq1 \end{subarray}} W(y|0) \left(\log\left(1 + L_{V}(y)\right)-\log2 + \log2\right)  \nonumber\\
\leq &1 - \displaystyle\sum_{y\in\cY} W(y|0) \log\left(\frac{V(y|0)}{q_V(y)}\right).
\end{align}
Similarly, the error probability given that the symbol $1$ is transmitted is upper bounded by 
\begin{align}\label{eq:Pe-ML-bound-1}
&P_{\textnormal{e, ML}}(W, V|1) := \displaystyle\sum_{\begin{subarray}{c}
y\colon\\L_{V}(y)<1 \end{subarray}} W(y|1) + \displaystyle\frac{1}{2}\displaystyle\sum_{\begin{subarray}{c}
y\colon\\L_{V}(y)=1 \end{subarray}}  W(y|1)  \nonumber\\
&\leq 1 - \displaystyle\sum_{y\in\cY} W(y|1) \log\left(\frac{V(y|1)}{q_V(y)}\right).
\end{align}
Thus, when both inputs are equally likely, we get
\begin{equation}\label{eq:Pe_I_mismatch_UB}
P_{\textnormal{e, ML}}(W, V) = \displaystyle\frac{1}{2}P_{\textnormal{e, ML}}(W, V|0) + \displaystyle\frac{1}{2}P_{\textnormal{e, ML}}(W, V|1) \leq \min\{1-I(W,V), 1\},
\end{equation}
where the inequality follows by \eqref{eq:Pe-ML-bound-0} and \eqref{eq:Pe-ML-bound-1} and the trivial upper bound $P_{\textnormal{e, ML}}(W, V) \leq 1$.
 \end{IEEEproof}
 
It is easy to see that when $V=W$, the parameter $I(W, V)$ recovers the symmetric capacity of the channel, i.e., $I(W) := I(W, W)$. For this reason, we call in this paper $I(W, V)$ as the ``mismatched mutual information".  Following this observation, the next fact we state shall not be surprising and is actually well-known in the mismatched decoding literature.
\begin{proposition}\label{prop:rem1-p4}
For any pairs of B-DMCs  $W\colon \fF_2\to\mathcal{Y}$ and $V\colon \fF_2\to\mathcal{Y}$, we have
\begin{equation}\label{eq:I(W,V)_ub}
I(W) \geq I(W, V). 
\end{equation}
\end{proposition}
\begin{IEEEproof}
To see that \eqref{eq:I(W,V)_ub} holds, one can let $W_{0} = W(y|0)$ and $W_{1} = W(y|1)$, and simply note that the difference can be written as:
\begin{multline}
I(W, V) - I(W) \\
= \displaystyle\frac{1}{2} \operatorname{Div}\left(W_{0} \| V_{0}\right) + \displaystyle\frac{1}{2} \operatorname{Div}\left(W_{1} \| V_{1}\right) 
- \operatorname{Div}\left(\displaystyle\frac{W_{0}+ W_{1}}{2} \bigg\|\displaystyle\frac{V_{0} + V_{1}}{2}\right) \geq 0,
\end{multline}
where $\operatorname{Div}(P_1\|P_2)$ denotes the Kullback–Leibler divergence and the inequality follows from the convexity of $\operatorname{Div}(P_1\|P_2)$ in the pair $(P_1, P_2)$ \cite{578869}. 
\end{IEEEproof}
Next, an important remark is in order.
 \begin{remark}\label{rem:rem1-p5p6} 
 By  the previous proposition, the bound $I(W, V)\leq 1$ necessarily holds. However, $I(W, V)$ can take on negative values, and in fact, it is unbounded from below. If $I(W, V) = -\infty$, then there must exist at least one output symbol $y\in\mathcal{Y}$ such that $V(y|x)=0$ but $W(y|x) >0$. 
\end{remark}

As a last point, we note that the following interpretation of the parameter appears in\cite[Proof of Corr. 6.10]{csiszar1981information}: 
\begin{equation}\label{eq:MMI-alternative-def}
I(W, V) = I(V) +  \displaystyle\lim_{\alpha\to 0} \displaystyle\frac{\partial}{\partial\alpha} I(P,\alpha W + (1-\alpha)V).
\end{equation}
This alternative interpretation of the parameter can be useful as follows. Let $\mathcal{\overline{W}}$ denote the convex closure of the set $\{W, V\}$ formed by our pair of mismatched channels. Now, if the channel $V$ is such that $V = \arg\min_{W'\in\mathcal{\overline{W}}}I(W')$, then, the lower bound $I(W,V) \geq I(V)$ holds. So, we obtain via \eqref{eq:MMI-alternative-def}, a sufficient condition for the non-negativity of $I(W,V)$.

\subsection{On the mismatched variational distance parameter}\label{subsec:pre-MVD}
We are not aware of any previous appearance of  the parameters $D(W, V)$ or $T(W, V)$ in the mismatched decoding literature. Nevertheless, we note that the quantity $T(W, V)$  can be interpreted as a ``mismatched variational distance'', since when $V = W$, we recover the variational distance between the distributions $W(y|0)$ and $W(y|1)$ given by
\begin{equation}
T(W) := T(W, W) = \frac{1}{2}\displaystyle\sum_{y\in\mathcal{Y}} \lvert W(y|0) - W(y|1)\rvert.
\end{equation}
Similarly, when $V = W$, we recover the following alternative distance measure between the square of the distributions $W(y|0)$ and $W(y|1)$:\footnote{One should not confuse the definition of $D(W)$ with the definition of the Hellinger distance between the distributions $W(y|0)$ and $W(y|1)$.}
\begin{equation}
D(W) := D(W, W) = \frac{1}{2}\displaystyle\sum_{y\in\mathcal{Y}} \sqrt{\lvert W(y|0)^2 - W(y|1)^2\rvert}.
\end{equation}
Furthermore, it is possible to show the following set of inequalities between the parameters.
\begin{proposition}\label{prop:rem2-p3}
For any pairs of B-DMCs  $W\colon \fF_2\to\mathcal{Y}$ and $V\colon \fF_2\to\mathcal{Y}$, we have
\begin{equation}\label{eq:ineq4}
D(W, V)^2 \leq T(W, V) \leq D(W, V),
\end{equation}
and 
\begin{equation}\label{eq:ineq5}
 T(W, V)^{k} \leq T_{k}(W, V) \leq T(W, V),
\end{equation}
for any $k\in\mathbb{N}\setminus\{0, 1\}$.
\end{proposition}
\begin{IEEEproof}
The inequalities in \eqref{eq:ineq4} follows by Jensen's inequality for concave functions and the inequality $\sqrt{x}\geq x$, for $x\in[0, 1]$, respectively. Similarly, the inequalities in \eqref{eq:ineq5} follows  by the inequality $x \geq x^k$, for $x\in[0, 1]$ and for any $k\in\mathbb{N}\setminus\{0, 1\}$, and Jensen's inequality for convex functions, respectively.
\end{IEEEproof}
Our second remark is in order.
 \begin{remark}\label{rem:rem2-p4} 
Since $\left|\Delta_{V}(Y)\right|\in[0, 1]$ and $\sqrt{\left|\Delta_{V}(Y)\right|}\in[0, 1]$, it follows that both $T(W, V)\in[0, 1]$ and $D(W, V)\in[0, 1]$ hold. Similarly, we have $T_{k}(W, V)\in[0, 1]$, for any $k\in\mathbb{N}\setminus\{0, 1\}$.
\end{remark}

We end this subsection by showing a general relation between the parameters $D(W, V)$ and $I(W, V)$.
\begin{proposition}\label{prop:D_I_coupling}
$I(W, V) \leq \displaystyle\frac{1}{\ln 2}D(W, V)$.
\end{proposition}
\begin{IEEEproof}
The result follows from the inequalities 
\begin{equation}
\log(1 + \Delta) \leq \sqrt{\lvert \Delta \rvert}/\ln 2 \quad \hbox{and} \quad \log(1 - \Delta) \leq \sqrt{\lvert \Delta \rvert}/\ln 2,
\end{equation}
which hold for $\Delta\in[-1, 1]$.
\end{IEEEproof}
 
\subsection{On the mismatched Bhattacharyya parameter}\label{subsec:pre-MZ}
Skipping the proof, we directly note that $Z(W, V)$ generates an upper bound on the mismatched ML decoding error probability.
\begin{proposition}\label{prop:rem3-p1}
For any pairs of B-DMCs  $W\colon \fF_2\to\mathcal{Y}$ and $V\colon \fF_2\to\mathcal{Y}$, we have
\begin{equation}\label{eq:Pe_bound_Z}
P_{\textnormal{e, ML}}(W, V) \leq Z(W,V). 
\end{equation}
\end{proposition}
We also note that when $V=W$, this parameter recovers the channel Bhattacharyya parameter, i.e., $Z(W, W) := Z(W)$. For this reason, we call in this paper $Z(W, V)$ as the ``mismatched Bhattacharyya parameter''.  However, while the Bhattacharyya parameter satisfies $Z(W)\in[0, 1]$, the mismatched version is unbounded from above as highlighted in the following remark.
 \begin{remark}\label{rem:rem3-p3} 
$Z(W, V)\geq 0$. In fact, $Z(W, V)=\infty$, whenever $I(W, V)=-\infty$.
\end{remark}

Finally, we look into how the parameters $D(W, V)$ and $Z(W, V)$ are related in general. 
\begin{proposition}\label{prop:D_Z_coupling}
$D(W,V) + Z(W, V) \geq 1$.
\end{proposition}
\begin{IEEEproof}
 One can easily prove the inequality by using the relation 
\begin{equation}\label{eq:delta_LR_relation}
 \Delta_V(y) = \tanh\left(\ln\left(\sqrt{\frac{1}{L_V(y)}}\right)\right),
\end{equation}
where $\tanh(\cdot)$ denotes the hyperbolic tangent function. To see this let $x>0$. First, note that we have $|\tanh(\ln{x})|=\tanh(|\ln{x}|)$, since $\tanh(-x)=-\tanh(x)$ holds for $x>0$. Next, we get $|\tanh(\ln{x})| \leq |\ln{x}|$ from the inequality $\tanh(x)\leq x$ which holds for $x>0$. Finally, by the inequality $\ln(x)\geq 1-1/x$, for $x>0$, we can write $|\ln{x}| \geq 1-1/x$ and $|\ln{x}| = |\ln{1/x}| \geq 1-x$. This proves that $|\tanh(\ln{x})|+x \geq 1$ holds for any $x>0$. Finally, using the identifications $x\leftarrow \sqrt{L_V(y)}$ and $x\leftarrow \sqrt{1/L_V(y)}$, respectively, in this latter inequality, the claim of the proposition is obtained.
\end{IEEEproof}
 
\section{Overview of the Main Results}\label{sec:Overview}
We introduce the main results of the paper in this section.  We state them in multiple subsections in a series of theorems--- Theorems \ref{thm:mismatched_polarization_D}, \ref{thm:rate_of_polarization_mismatched_setting}, \ref{thm:coding}, \ref{thm:mismatch_family_of_LBs}, and \ref{thm:rate_of_polarization_new} --- and we discuss key ideas involved in their proofs. The complete proofs are carried out in Section \ref{sec:polar_mismatched_capacity}.

For the theorem statements, we need to define the stochastic processes corresponding to the introduced mismatched channel parameters. For $n\in\mathbb{N}$, we let $I_n(W, V) := I(W_n, V_n)$, $D_n(W, V):= D(W_n, V_n)$, $T_n(W, V) := T(W_n, V_n)$, and $Z_n(W, V) := Z(W_n, V_n)$ denote mismatched channel parameter processes associated to the two channel polarization processes $W_n$ and $V_n$ defined by the recursion in \eqref{eq:def_pol_process}. The matched versions of these processes are simply obtained by dropping their second argument, i.e., $I_n(W):=I_n(W, W)$, $D_n(W):=D_n(W, W)$, $T_n(W):=T_n(W, W)$, and $Z_n(W):=Z_n(W, W)$. Finally, we use the notation  $I_{\infty}(W, V):= \lim_{n\to\infty} I_n(W, V)$ to define the limiting random variable in the sense of a.s. convergence of the process, provided that the limit exists, and denote similarly by $D_{\infty}(W, V), T_{\infty}(W, V)$, and $Z_{\infty}(W, V)$ the limits of the processes $D_n(W, V)$, $T_n(W, V)$, and $Z_n(W, V)$, respectively, in the sense of a.s. convergence, provided that the limits exist.

\subsection{Channel Polarization in the mismatched setting}
In the next theorem, we characterize the polarization phenomenon in the mismatched setting.
\begin{theorem}\label{thm:mismatched_polarization_D}
Let $N=2^n$ with $n\in\mathbb{N}$. For any two B-DMCs $W:\mathbb{F}_2\to\mathcal{Y}$ and $V:\mathbb{F}_2\to\mathcal{Y}$ such that $I(W, V) > -\infty$,  the mismatched pair of channels $(W, V)$ polarize in the sense that,  for all $\gamma\in(0, 1/2)$, we have
  \begin{equation} \label{eq:polarization_Pe_1} 
\frac{1}{N} \hspace{2mm} \# \{i\in\{1, \ldots, N\}\ssep P_{\textnormal{e, ML}}(W_{N}^{(i)}, V_{N}^{(i)}) \in (\gamma, 1/2- \gamma)\} \xrightarrow[N\to\infty]{} 0, 
\end{equation}
such that
\begin{equation} \label{eq:polarization_Pe_2} 
\frac{1}{N} \hspace{2mm} \# \{i\in\{1, \ldots, N\}\ssep P_{\textnormal{e, ML}}(W_{N}^{(i)}, V_{N}^{(i)}) \in  (0, \gamma]\} \xrightarrow[N\to\infty]{}  C_\mathrm{P}(W,V) , 
\end{equation}
\begin{equation} \label{eq:polarization_Pe_3} 
\frac{1}{N} \hspace{2mm} \# \{i\in\{1, \ldots, N\}\ssep P_{\textnormal{e, ML}}(W_{N}^{(i)}, V_{N}^{(i)}) \in [1/2-\gamma, 1/2)\} \xrightarrow[N\to\infty]{} 1-C_\mathrm{P}(W,V) , 
\end{equation}
where $C_\mathrm{P}(W,V):= \Prob[D_{\infty}(W, V)= 1]$. 
\end{theorem}
Theorem \ref{thm:mismatched_polarization_D} can be seen as the mismatched counterpart of Theorem \ref{thm:polarization}. In fact, one can check that for $V = W$, we have $\Prob[D_{\infty}(W)= 1] = \Prob[I_{\infty}(W)=1]$.\footnote{The relation $\Prob[D_{\infty}(W)= 1] \geq \Prob[I_{\infty}(W)=1]$ can be verified by using the relation given in Proposition \ref{prop:D_I_coupling} stated in Section \ref{subsec:pre-MVD} and noting that $D_n\in [0, 1]$, for all $n\in\mathbb{N}$.} Thus, the statement of the theorem is aligned with the original result of Theorem \ref{thm:polarization} where $\Prob[I_{\infty}(W)=1] = I(W) =  C_\mathrm{P}(W, W) = \Prob[D_{\infty}(W)= 1]$.  The proof of this theorem will be carried out in Section \ref{subsec:proof_mismatched_polarization_D}. In the outline below, we highlight the main conceptual steps involved in this proof.
\begin{IEEEproof}[Proof Outline]
The proof of the theorem will rely on the convergence properties of the mismatched processes $D_n(W, V)$ and $I_n(W, V)$, for $n\in\mathbb{N}$, investigated in Section \ref{subsec:proof_mismatched_polarization_D}. Regarding the process $D_n(W, V)$, we will prove in Proposition \ref{prop:D_n_supermartingale} that $D_n(W, V)$ is a bounded super-martingale which converges a.s. to a limiting random variable $D_\infty(W, V)$ taking values in the set $\{0, 1\}$ a.s. As a consequence, the following polarization result will readily follow: for all $\gamma\in(0, 1)$, 
  \begin{equation} \label{eq:polarization_D_1} 
\frac{1}{N} \hspace{2mm} \# \{i\in\{1, \ldots, N\}\ssep D(W_{N}^{(i)}, V_{N}^{(i)}) \in (\gamma, 1- \gamma)\} \xrightarrow[N\to\infty]{} 0, 
\end{equation}
such that
\begin{equation} \label{eq:polarization_D_2} 
\frac{1}{N} \hspace{2mm} \# \{i\in\{1, \ldots, N\}\ssep D(W_{N}^{(i)}, V_{N}^{(i)}) \in  [1-\gamma, 1)\} \xrightarrow[N\to\infty]{}  C_\mathrm{P}(W,V) , 
\end{equation}
\begin{equation} \label{eq:polarization_D_3} 
\frac{1}{N} \hspace{2mm} \# \{i\in\{1, \ldots, N\}\ssep D(W_{N}^{(i)}, V_{N}^{(i)}) \in (0, \gamma]\} \xrightarrow[N\to\infty]{} 1-C_\mathrm{P}(W,V).
\end{equation}

Although, the above three equations look similar to \eqref{eq:polarization_Pe_1}, \eqref{eq:polarization_Pe_2}, and \eqref{eq:polarization_Pe_3}, respectively, it is not clear what is the operational meaning of this particular polarization phenomenon. To clarify this and prove the theorem, the implications of the convergences described in \eqref{eq:polarization_D_2} and \eqref{eq:polarization_D_3} on the convergence properties of the error probability process $P_{\textnormal{e, ML}}(W_n, V_n)$ will be investigated. Part of this investigation will make use of the convergence properties of the process $I_n(W, V)$ in addition to the convergence properties of the process $D_n(W, V)$. Regarding the process $I_n(W, V)$, we will prove in Proposition \ref{prop:I_n_martingale} that the mismatched mutual information process $I_n(W, V)$ is, as its matched counterpart, a martingale process. Ultimately, the provided analysis in Section \ref{subsec:proof_mismatched_polarization_D} will reveal the following operational meanings about the convergence points of the process $D_n(W, V)$:
\begin{itemize}
\item We will conclude via Proposition \ref{prop:Delta_zero_submartingale} that  $D_\infty(W, V) = 0$ implies $\lim_{n\to\infty}P_{\textnormal{e, ML}}(W_n, V_n) =1/2$ a.s. Hence, after a long sequence of polar transformations, the input bits whose indices belong to the fraction in \eqref{eq:polarization_D_3} see almost completely noisy synthetic channels $W_N^{(i)}$ over which the corresponding genie aided mismatched ML decoders with respect to the channels $V_N^{(i)}$ fail with high probability. Thus, we will have shown that these same indices do also belong to the fraction in \eqref{eq:polarization_Pe_3}.
\item We will conclude via Proposition \ref{prop:Delta-convergence-1} that $D_\infty(W, V) = 1$ implies $\lim_{n\to\infty} P_{\textnormal{e, ML}}(W_n, V_n)= 0$ a.s., as long as $I(W, V) > -\infty$ holds. Hence, given the latter condition is satisfied, after a long sequence of polar transformations,  the input bits whose indices belong to the fraction in \eqref{eq:polarization_D_2} will see almost perfect synthetic channels $W_N^{(i)}$ where uncoded transmission results in a vanishing error probability under mismatched ML decoding with respect to the channels $V_N^{(i)}$. So this time, we will have shown that the indices do belong to the fraction in \eqref{eq:polarization_Pe_2}.
\end{itemize} 
\end{IEEEproof} \vspace*{-2mm}

\subsection{Coding Theorem with mismatched polar decoding}\label{subsec:coding-theorem}
We saw in the previous subsection that a polarization phenomenon also holds in the mismatched setting. Naturally, our goal is to explore the implications of this new phenomenon from a channel coding perspective. Theorem \ref{thm:mismatched_polarization_D}, in which we observed the polarization result in terms of the quantities $P_{\textnormal{e, ML}}(W_N^{(i)}, V_N^{(i)})$, for $i\in\{1, \ldots, N\}$, suggests to consider polar code constructions with the following choice of information sets:
\begin{equation}\label{eq:info_set_mismatch}
\mathcal{A}_{N, \gamma}(W, V) := \big\{i\in\{1, \ldots, N\}\ssep P_{\textnormal{e, ML}}(W_N^{(i)}, V_N^{(i)})\leq \gamma\big\},
\end{equation} 
where $N = 2^n$ with $n\in\mathbb{N}$ and $\gamma\in(0, 1/2)$ is a desired threshold. We know from \eqref{eq:polarization_Pe_2} that while the synthetic channels polarize as $N$ grows through powers of two, the
fraction $|\mathcal{A}_{N, \gamma}(W, V)|/N$ tends to $C_P(W, V)\in[0, 1]$, for any $\gamma\in(0, 1/2)$. By the definition of $P_{\textnormal{e, ML}}(W_N^{(i)}, V_N^{(i)})$, we know that over any synthetic channel $W_{N}^{(i)}$ such that $i \in \mathcal{A}_{N, \gamma}(W, V)$,  uncoded transmission will result in a vanishing decoding error probability even if the receiver performs ML decoding with respect to the law of the possibly mismatched channel $V_{N}^{(i)}$.
Moreover, for any $i \notin \mathcal{A}_{N, \gamma}(W, V)$, transmission over $W_{N}^{(i)}$ will fail with high probability using such a receiver. Although each sequence $P_{\textnormal{e, ML}}(W_N^{(i)}, V_N^{(i)})$ vanishes for every $i\in\mathcal{A}_{N, \gamma}(W, V)$ when $N\to\infty$ and $\gamma\to0$, whether  the block decoding error probability sequence ${P_\textnormal{e}}(W, V, \mathcal{A}_{N, \gamma}(W, V))$ of the mismatched polar successive cancellation decoder vanishes or not is still subtle. As in the matched setting, we need to investigate the rate of polarization, i.e., the speed with which the variable $\gamma$ in \eqref{eq:info_set_mismatch} can be made to approach zero as a function of the blocklength $N$. The  next theorem states our result regarding this rate of convergence.

\begin{theorem}\label{thm:rate_of_polarization_mismatched_setting}
For any two B-DMCs $W:\mathbb{F}_2\to\mathcal{Y}$ and $V:\mathbb{F}_2\to\mathcal{Y}$, any fixed rate $R<C_\mathrm{P}(W, V)$, and constant $\beta\in(0,1/2)$, there exists a sequence of sets $\mathcal{A}_N \subseteq \{1, \ldots, N\}$, with $N=2^n$ for $n\in\mathbb{N}$, such that $|\mathcal{A}_N|\geq NR$, and 
\begin{equation}
\displaystyle\sum_{i\in\mathcal{A}_N}P_{\textnormal{e, ML}}(W_{N}^{(i)}, V_{N}^{(i)}) = o(2^{-N^\beta}).
\end{equation}
\end{theorem}
The proof of this theorem will be carried out in Section \ref{subsec:proof_rate_of_polarization_mismatched_setting}. Below, we present the key ideas involved in this proof.
\begin{IEEEproof}[Proof Outline]
The idea of the proof will be to extend the rate of convergence result stated in the first part of Theorem \ref{thm:rate_of_polarization_Qn}, i.e., the claim in  \eqref{eq:Qn_convergence_1}, to the mismatched setting. This will be achieved by studying the evolution of the process $Z_n(W, V)$ under the polar transform and by taking a closer look at the proof of Theorem \ref{thm:rate_of_polarization_Qn} carried out in \cite{5205856}. More specifically, we will show  in Section \ref{subsec:proof_rate_of_polarization_mismatched_setting}  that while the mismatched Bhattacharyya parameter process $Z_n(W, V)$ readily satisfies the condition $(c.2)$, it does not fully satisfy the conditions $(c.1)$ and $(c.3)$. Nevertheless, we will see that the proof of Theorem \ref{thm:rate_of_polarization_Qn} does not require the conditions $(c.1)$ and $(c.3)$ to fully hold, and in fact, it is possible to adapt these conditions to the characteristics of the mismatched Bhattacharyya parameter process $Z_n(W, V)$, and obtain the desired result.
 \end{IEEEproof}

The following coding theorem follows as a corollary to Theorem \ref{thm:mismatched_polarization_D} and Theorem \ref{thm:rate_of_polarization_mismatched_setting}. 
\begin{theorem}\label{thm:coding}
For any two B-DMCs $W:\mathbb{F}_2\to\mathcal{Y}$ and $V:\mathbb{F}_2\to\mathcal{Y}$ such that $I(W, V) > -\infty$,  polar coding with mismatched polar successive cancellation decoding at any fixed rate $R < C_\mathrm{P}(W, V)$ satisfies 
\begin{equation}\label{eq:coding_thm}
P_\mathrm{e}(W, V, \mathcal{A}_{N}^{*}) = o(2^{-N^\beta}),
\end{equation} 
for any constant $\beta\in(0, 1/2)$, threshold $\gamma=o(2^{-N^\beta})$, and any information subset $\mathcal{A}_{N}^{*} \subseteq \mathcal{A}_{N, \gamma}(W, V)$ of size $\lfloor RN\rfloor$ selected according to the definition in
\eqref{eq:info_set_mismatch}.   
\end{theorem}

Theorem \ref{thm:coding}, in view of Theorem \ref{thm:mismatched_polarization_D}, reveals that the quantity $C_\mathrm{P}(W, V)$ corresponds to the transmission capacity of polar coding over the channel $W$ when the outputs are decoded with a mismatched polar successive cancellation decoder operating with the metric of a B-DMC $V$ not necessarily matched to $W$. We call $C_\mathrm{P}(W, V)$ as the \textit{polar mismatched capacity}. 

\begin{remark}\normalfont
In Theorem \ref{thm:rate_of_polarization_mismatched_setting}, we did not introduce a mismatched counterpart for the second part of Theorem \ref{thm:rate_of_polarization_Qn}, i.e., we did not show that a result of the form \eqref{eq:Qn_convergence_2} holds for any $\beta>1/2$.
However, we know that if $V=W$, then \eqref{eq:Qn_convergence_2} is true by \cite[Thm. 3]{5205856}, since the Bhattacharyya parameter process $Z_n(W)$ satisfies the conditions of the theorem. As by the hypothesis of Theorem \ref{thm:rate_of_polarization_mismatched_setting}, the channels $W: \mathbb{F}_2\to\mathcal{Y}$ and $V:\mathbb{F}_2\to\mathcal{Y}$ can be any two B-DMC satisfying the condition $I(W, V) > -\infty$, including the case $V=W$, we conclude that, under the general assumption of the theorem, we cannot improve the rate of convergence result stated in \eqref{eq:fast_convergence_Z} beyond the interval $\beta\in(0, 1/2)$.
\end{remark}

To achieve a positive rate in the mismatched setting, we note that $C_\mathrm{P}(W,V) > 0$ must hold. The next theorem complements the previous coding theorem by providing sufficient conditions for the strict positivity of the polar mismatched capacity. 
\begin{theorem}\label{thm:mismatch_family_of_LBs}
For any two B-DMCs $W:\mathbb{F}_2\to\mathcal{Y}$ and $V:\mathbb{F}_2\to\mathcal{Y}$, the following improving family of lower bounds on $C_\mathrm{P}(W,V)$ holds for all  $N = 2^n$ with $n\in\mathbb{N}$:
\begin{equation}\label{eq:mismatch_family_of_LBs}
C_\mathrm{P}(W,V) \geq \displaystyle\frac{1}{N} \displaystyle\sum_{i=1}^{N} \big|I(W_N^{(i)}, V_N^{(i)})\big|^+,
\end{equation}
where $\big|I(W_N^{(i)}, V_N^{(i)})\big|^+ = \max\{I(W_N^{(i)}, V_N^{(i)}), 0\}$. Furthermore, the family of lower bounds in \eqref{eq:mismatch_family_of_LBs} is indeed asymptotically tight.
\end{theorem}
From Theorem \ref{thm:mismatch_family_of_LBs}, we can see that the positivity of the mismatched mutual information, i.e., $I(W, V) > 0$, is a sufficient condition for $C_\mathrm{P}(W,V) > 0$.  In that respect, the discussion following the alternative representation of $I(W, V)$ given in \eqref{eq:MMI-alternative-def} is also relevant here, as it leads to a particular channel condition for the non-negativity of $I(W, V)$. The proof of the theorem will be completed in Section \ref{subsec:proof_mismatch_family_of_LBs}. We proceed with a brief discussion on the proof idea.
\begin{IEEEproof}[Proof Outline]
The proof of the theorem will heavily rely on the following technical result:
\begin{equation}
\Prob[D_\infty(W, V) = 1] \geq I(W, V).
\end{equation}
We will re-state this result in Proposition \ref{prop:LB_C_P} in Section \ref{subsec:proof_mismatch_family_of_LBs}, where it will also be proved. 
\end{IEEEproof}

To get some intuition on the family of lower bounds given in \eqref{eq:mismatch_family_of_LBs},  we now look at few examples of classes of mismatched channel pairs $(W, V)$.
\begin{example}\normalfont
Let $W$ and $V$ be two BSCs of crossover probabilities $p\in[0, 1]$ and $q\in[0,1]$, respectively. In this case, we have
\begin{equation}
I(W,V ) = 1- h_2(p)-D_2(p \Vert q),
\end{equation}
where $h_2(\cdot)$ denotes the binary entropy function and $D_2(p \Vert q) := \operatorname{Div}(W(\cdot|0)\Vert V(\cdot|0))$.  Thus, $I(W)-I(W,V)=D_2(p \Vert q)$. 
\end{example}

\begin{example}\normalfont
Let $W$ and $V$ be two binary erasure channels (BECs) of erasure probabilities $p\in[0, 1]$ and $q\in[0,1]$, respectively. Then, it turns out that
\begin{equation}
I(W,V ) = I(W) = 1-p.
\end{equation}
Thus, for pairs of BECs, the polar mismatched capacity equals the matched capacity of the communication channel, i.e., $C_\mathrm{P}(W,V) = I(W)$. 
\end{example}

\begin{example}\normalfont \cite{6970855} \label{ex:Example}
Let $W$ be a BSC of crossover probability $p\in(0, 0.5)$ and $V$ be the BSC of crossover probability $1-p$. Suppose that we apply the polar transform to synthesize the channels $W^+$, $W^-$ and $V^+$, $V^-$. It is known that after applying the minus polar transform to a BSC of crossover probability $\alpha\in[0, 1]$, 
the synthesized channel is also a BSC, and with crossover probability $2\alpha(1-\alpha)$. So, both $W^-$ and $V^-$ are the same BSC with crossover probability $p^- = 2p(1-p)$. It is also easy to see that while $V^+\neq W^+$, one has $V^{+-}=W^{+-}$, and indeed,
$V^{++-}=W^{++-}$, \dots.  Consequently, for any sequence $s^n\in\{-, +\}^n$ of polar transforms, $V^{s^n}=W^{s^n}$, 
except when $s^n=+\dots+$. In this case $C_\mathrm{P}(W,V)=I(W)$, and we thus have $\mathcal{A}_{N, \gamma}(W, V) = \mathcal{A}_{N, \gamma}(W, W)$, for any $\gamma > 0$. 
\end{example}

\begin{example}\normalfont
Let $W$ be a BSC of crossover probability $p\in[0, 1]$ and $V$ be a BEC of erasure probability $q\in[0,1]$. Then, we have $I(W, V) = -\infty$. Going one-step further,  since the channel $W^-$ is a BSC and the channel $V^-$ is a BEC, we get similarly $I(W^-, V^-) = -\infty$. On the other hand, a simple calculation shows that we also have $I(W^-, V^-) = -\infty$. Furthermore, since it is well-known that all the channels synthesized from the BEC $V$ are also BECs \cite{1669570} and the channel W is a BSC, we can always find an output symbol $y^*$ such that $W^{s^n}(y^*|0) >0$, but  $V^{s^n}(y^*|0) = 0$, for any $s^n \in \{-, +\}^n$.  As pointed out in Remark \ref{rem:rem1-p5p6}, it follows by this observation that $I(W^{s^n}, V^{s^n}) = -\infty$ holds, for all $s^n\in \{-, +\}^n$.
\end{example}

\subsection{From mismatched to matched setting}
While the convergence speed of the Bhattacharyya parameter process $Z_n(W)$ to the limiting value $Z_\infty(W)=0$  as a function of the blocklength has been investigated in \cite{5205856}, it was pointed out in \cite{tal-2016} that no such result has been given for the speed of convergence to the limiting value $Z_\infty(W)=1$.  In the next theorem, we show that a similar rate of convergence result holds as well for the latter convergence.
 \begin{theorem}\label{thm:rate_of_polarization_new}
For any B-DMC $W:\mathbb{F}_2\to\mathcal{Y}$, any fixed rate $R<I(W)$, and constant $\beta\in(0, 1/2)$, there exists a sequence of sets $\mathcal{A}_N \subseteq \{1, \ldots, N\}$, for $N=2^n$ with $n\in\mathbb{N}$, such that $|\mathcal{A}_N|\geq NR$ and 
\begin{equation}\label{eq:rate_of_polarization_new}
Z(W_{N}^{(i)}) \geq 1 - o(2^{-N^\beta}),
\end{equation}
for any $i\notin\mathcal{A}_N$. Conversely, if $R>0$ and $\beta>1/2$, then for any sequence of  sets $\mathcal{A}_N \subseteq \{1, \ldots, N\}$ such that $|\mathcal{A}_N|\geq NR$, we have 
\begin{equation}\label{eq:rate_of_polarization_new_2}
\min\bigg\{Z(W_{N}^{(i)}): i\notin\mathcal{A}_N\bigg\} = 1-\omega(2^{-N^\beta}).
\end{equation} 
\end{theorem}
The theorem will be proved in Section \ref{subsec:proof_rate_of_polarization_new}. We proceed with the discussion of the proof idea.
\begin{proofoutline}
We will first argue in Proposition  \ref{prop:rate_of_convergence_D_to_zero} that ``\textit{whenever $D_n(W, V)$ converges to zero, this convergence is almost surely fast}'', i.e., $\lim_{n\to\infty}\Prob[D_n(W, V) < 2^{-2^{n\beta}}] \to 1 - I(W)$, for any $\beta\in(0, 1/2)$. This argument will follow as a corollary to the first part of Theorem \ref{thm:rate_of_polarization_Qn} in view of the properties we derive for the process $D_n(W, V)$ in Section \ref{subsec:proof_mismatched_polarization_D}. Then, to show \eqref{eq:rate_of_polarization_new}, we will explore the coupling between the limiting random variables $D_\infty(W)$ and $Z_\infty(W)$ via Proposition \ref{prop:D_Z_coupling} derived in Section \ref{subsec:pre-MZ}. The claim in \eqref{eq:rate_of_polarization_new_2} will be proved similarly by using the second part of Theorem \ref{thm:rate_of_polarization_Qn}. This time, we will study some properties of the process $T_n(W)$, and the relation between the limits $Z_\infty(W)$ and $T_\infty(W)$. The existence of this latter limit will follow from the proof of Proposition \ref{prop:Delta-convergence-prob}.
\end{proofoutline}

\section{Proofs of the Theorems}\label{sec:polar_mismatched_capacity}
In this section, we prove Theorems \ref{thm:mismatched_polarization_D}, \ref{thm:rate_of_polarization_mismatched_setting}, \ref{thm:mismatch_family_of_LBs}, and \ref{thm:rate_of_polarization_new}. As previously, we assume that $W:\mathbb{F}_2\to\mathcal{Y}$ and $V:\mathbb{F}_2\to\mathcal{Y}$ denote a pair of mismatched B-DMCs. 

\subsection{Proof of Theorem \ref{thm:mismatched_polarization_D}}\label{subsec:proof_mismatched_polarization_D}
To prove the theorem, we will study the local transformation of the mismatched parameters  under the polar transform  and the convergence properties of the corresponding stochastic processes. We start the analysis with the parameter $D(W, V)$. 

\begin{proposition}\label{prop:D_evolution}
The polar transform defined in \eqref{eq:def_pol_minus} and \eqref{eq:def_pol_plus} exhibits the following properties:
\begin{align}
\label{eq:D_minus}&D(W^{-}, V^{-}) = D(W, V)^2, \\
\label{eq:D_plus}&D(W^{+}, V^{+}) \leq 2D(W, V) - D(W, V)^2. 
\end{align}
\end{proposition}
\begin{IEEEproof} 
It is easy to show that 
\begin{align}
\label{eq:d_minus}\Delta_{V^{-}}(Y_1Y_2) &= \Delta_{V}(Y_1)\Delta_{V}(Y_2), \\
\label{eq:d_plus}\Delta_{V^{+}}(Y_1Y_2U_1) &= \displaystyle\frac{\Delta_{V}(Y_1) + (-1)^{U_1} \Delta_{V}(Y_2)}{1 + (-1)^{U_1}\Delta_{V}(Y_1)\Delta_{V}(Y_2)},
\end{align}
where $Y_1\sim q_W(y_1)$, $Y_2\sim q_W(y_2)$, and the conditional distribution of $U_1$ given $Y_1$ and $Y_2$ equals $p(u_1|y_1y_2) = \frac{1}{2}\left(1 + (-1)^{u_1}\Delta_{W}(y_1)\Delta_{W}(y_2)\right)$. So, we have 
\begin{equation}
D(W^-, V^-) = \Expt\left[\sqrt{|\Delta_{V^-}(Y_1Y_2)|}\right] = \Expt\left[\sqrt{|\Delta_{V}(Y_1)|}\right]\Expt\left[\sqrt{|\Delta_{V}(Y_2)|}\right] = D(W, V)^2,
\end{equation}
and
\begin{multline}
D(W^+, V^+) =  \Expt\left[\sqrt{\left|\Delta_{V^+}(Y_1Y_2U_1)\right|}\right] \\
\stackrel{(a)}{\leq} \Expt\left[\sqrt{|\Delta_{V}(Y_1)|}\right]+\Expt\left[\sqrt{|\Delta_{V}(Y_2)|}\right] - \Expt\left[\sqrt{|\Delta_{V}(Y_1)|}\right]\Expt\left[\sqrt{|\Delta_{V}(Y_2)|}\right] \\
 = 2D(W, V) - D(W, V)^2,
\end{multline}
where the inequality $(a)$ holds by Lemma \ref{lem:delta_plus} proved in the Appendix.
\end{IEEEproof}
In the light of the local behavior uncovered by the previous proposition, we get the following result on the process $D_n(W, V)$.
\begin{proposition}\label{prop:D_n_supermartingale}
The process $D_n(W, V)$, for $n\in\mathbb{N}$, is a bounded super-martingale which converges a.s. toward a limiting random variable $D_\infty(W, V)$ taking values in $\{0, 1\}$. 
\end{proposition}
\begin{IEEEproof} 
By the observation in Remark \ref{rem:rem2-p4},  $D_n(W, V)\in[0, 1]$ holds for all $n\in\mathbb{N}$,  and thus, the process is bounded. The claim that the process is a super-martingale follows by Proposition \ref{prop:D_evolution} and the recursive structure of the transformation, i.e., by the observation that the recursive application of the polar transform  leads to a process such that:
\begin{align}
\label{eq:Dn_minus}&D_{n+1}(W, V) = D_n(W, V)^2, &\quad \hbox{when } B_{n+1} = 0,\\
\label{eq:Dn_plus}&D_{n+1}(W, V) \leq 2D_n(W, V) - D_n(W, V)^2, &\quad \hbox{when } B_{n+1} = 1,
\end{align}
where $B_n$ is as defined in \eqref{eq:def_pol_process}. By general results on bounded martingales, the process  $D_n(W, V)$ is uniformly integrable, and hence it converges a.s. and in $\mathcal{L}_1$ to a limiting random variable $D_\infty(W, V)$ such that $\Expt[\lvert D_n(W, V) - D_\infty(W, V)\rvert] \to 0$ as $n\to\infty$, see for instance \cite[Chap. 11]{Martingales:Book}. This implies that we have $\Expt[|D_{n+1}(W, V)-D_n(W, V)|]\to 0$. Since we can lower bound this expectation as 
$\Expt[|D_{n+1}(W, V)-D_n(W)|] \geq 1/2\Expt[D_n(W, V)(1-D_n(W, V))] \geq 0$ by using the squaring property  of the minus transform given in \eqref{eq:Dn_minus}, it follows that $\Expt[D_\infty(W, V)(1-D_\infty(W, V))]\to 0$.  Thus, $D_\infty(W, V)\in\{0, 1\}$ holds. A similar argument has been used in the proof of \cite[Prop. 9]{1669570}.
\end{IEEEproof}
Recall that in Section \ref{sec:pre-analysis}, the channel polarization process has been suitably defined on the probability space $(\Omega, \mathscr{F}_{\infty}, \Prob)$. In order to distinguish the paths $w = (w_1, w_2, \ldots)\in\Omega$ of the process according to their convergence properties, we define the following two sets:
\begin{equation}\label{eq:D0-set}
\mathcal{D}_0 := \{w\in\Omega: \lim_{n\to\infty} D(W^{s^n(w)}, V^{s^n(w)}) = 0\},
\end{equation}
and 
\begin{equation}\label{eq:D1-set}
\mathcal{D}_1 := \{w\in\Omega: \lim_{n\to\infty} D(W^{s^n(w)}, V^{s^n(w)}) = 1\},
\end{equation}
where we abused notation to define the function $s^n(w):= (s_1, \ldots s_n)$ via the mapping
 \begin{equation}\label{eq:sign-sequence-s}
s_{j} =  \left\{\begin{array}{ll}
                           -, & \hbox{if}  \hspace{3mm} w_j=0 \\
		     +, &\hbox{if}  \hspace{3mm}  w_j=1
                                \end{array} \right. .
\end{equation} 
By Proposition \ref{prop:D_n_supermartingale}, we know that $\Prob[\mathcal{D}_0 \cup \mathcal{D}_1]= 1$. To recover the index of a synthetic channel on a path $w\in\Omega$, we similarly abuse notation by defining the function $i^n(w) :=  1 + \sum_{j=1}^{n}w_j2^{n-j}$. 

\begin{proposition}\label{prop:Delta-convergence-prob}
For any $w\in\Omega$, the process $\lvert\Delta_{V^{s^n(w)}}\left(Y^*\right)\rvert$, for $n\in\mathbb{N}$, converges in probability under the distribution $q_{W^{s^n(w)}}(\cdot)$, and the sequence $\lvert\Delta_{V^{s^n(w)}}\left(Y^*\right)\rvert$, for $n\in\mathbb{N}$, is uniformly integrable. Furthermore, for any $w\in\mathcal{D}_0$, the process  $\lvert\Delta_{V^{s^n(w)}}\left(Y^*\right)\rvert$ converges in probability toward $0$ under the distribution $q_{W^{s^n(w)}}(\cdot)$, i.e., for all $\epsilon > 0$,
\begin{equation}\label{eq:D0-convergence-0}
q_{W^{s^n(w)}}\left(\left\lbrace y^*\in\mathcal{Y}_1^{2^n}{\mathbb{F}_2}^{i^n(w)-1}: \lvert\Delta_{V^{s^n(w)}}(y^*)\rvert \geq \epsilon \right\rbrace \right) \xrightarrow[n\to \infty]{} 0.
\end{equation}
For any $w\in\mathcal{D}_1$, the process  $\lvert\Delta_{V^{s^n(w)}}\left(Y^*\right)\rvert$ converges in probability toward $1$ under the distribution $q_{W^{s^n(w)}}(\cdot)$, i.e., for all $\epsilon > 0$,
\begin{equation}\label{eq:D1-convergence-1}
q_{W^{s^n(w)}}\left(\left\lbrace y^*\in\mathcal{Y}_1^{2^n}{\mathbb{F}_2}^{i^n(w)-1}: \lvert\Delta_{V^{s^n(w)}}(y^*)\rvert \leq 1-\epsilon \right\rbrace \right) \xrightarrow[n\to \infty]{} 0.
\end{equation}
The above assertions also hold for the family of processes $\lvert\Delta_{V^{s^n(w)}}\left(Y^*\right)\rvert^{k}$, for any $k\in\mathbb{N}\setminus\{0, 1\}$.
\end{proposition}
\begin{IEEEproof} 
By \eqref{eq:ineq4} stated in Proposition \ref{prop:rem2-p3}, we note that 
\begin{equation}
D(W^{s^n(w)}, V^{s^n(w)})\leq T(W^{s^n(w)}, V^{s^n(w)}) \leq D(W^{s^n(w)}, V^{s^n(w)})^2
\end{equation}
 holds, for any $w\in\Omega$ and for all $n\in\mathbb{N}$. Thus, taking limit superior and limit inferior, 
\begin{equation}
D_{\infty}(W, V)\leq \displaystyle\liminf_{n\to\infty}T(W^{s^n(w)}, V^{s^n(w)}) \leq \displaystyle\limsup_{n\to\infty}T(W^{s^n(w)}, V^{s^n(w)}) \leq D_{\infty}(W, V)^2
\end{equation}
holds a.s. by the squeeze theorem. We conclude by Proposition \ref{prop:D_n_supermartingale} that the process $T_n(W, V)$ converges a.s to a limiting random variable $T_{\infty}(W, V)$ which satisfies $T_{\infty}(W, V) = D_{\infty}(W, V)\in\{0, 1\}$ a.s. 
By this assertion, we can write
\begin{equation}\label{eq:T_n_0_case}
T(W^{s^n(w)}, V^{s^n(w)})  = \Expt\left[\big\lvert\Delta_{V^{s^n(w)}}\left(Y^*\right)\big\rvert\right] \xrightarrow[n\to \infty]{} \{0, 1\},
\end{equation}
for any  $w\in\Omega$, where the expectation is with respect to the distribution $q_{W^{s^n(w)}}(\cdot)$. The convergence in \eqref{eq:T_n_0_case}, in turn, implies that the process $\lvert\Delta_{V^{s^n}}\left(Y^*\right)\rvert$ converges in probability, and the sequence of random variables is uniformly integrable, see for instance \cite[Sec. 13.7]{Martingales:Book}. Consequently, for $w\in\mathcal{D}_0$,  the process $\lvert\Delta_{V^{s^n}}\left(Y^*\right)\rvert$ converges toward a random variable taking the value $0$, and for $w\in\mathcal{D}_1$, it converges toward a random variable taking the value $1$. This proves the claims in \eqref{eq:D0-convergence-0} and \eqref{eq:D1-convergence-1}. The final claim of the lemma can be proved via similar argument by using the relation \eqref{eq:ineq5} stated in Proposition \ref{prop:rem2-p3}.
\end{IEEEproof}
An immediate consequence of Proposition \ref{prop:Delta-convergence-prob} is that the process  $\Delta_{V^{s^n(w)}}\left(Y^*\right)$ measured under the distribution $q_{W^{s^n(w)}}(\cdot)$ converges in probability toward a $\{-1, 0, 1\}$-valued random variable, for all $w\in\Omega$  except possibly for a set of points of measure zero. 
To prove Theorem \ref{thm:mismatched_polarization_D}, however, we need more refined versions of the convergences stated in \eqref{eq:D0-convergence-0} and \eqref{eq:D1-convergence-1}, for the paths $w\in \mathcal{D}_0$ and $w\in \mathcal{D}_1$, respectively. So, we investigate further the convergence properties of the process $\Delta_{V^{s^n(w)}}\left(Y^*\right)$ for these paths.

\begin{proposition}\label{prop:Delta_zero_submartingale}
The process $q_{W^{s^n(w)}}\left(\left\lbrace y^*\in\mathcal{Y}_1^{2^n}{\mathbb{F}_2}^{i^n(w)-1}: \Delta_{V^{s^n(w)}}(y^*) = 0 \right\rbrace \right)$ is a bounded sub-martingale process taking values in $[0, 1]$ and the process converges a.s. toward a limiting random variable taking values in $\{0, 1\}$. Furthermore, 
\begin{equation} \label{eq:Delta_zero_submartingale_D0}
q_{W^{s^n(w)}}\left(\left\lbrace y^*\in\mathcal{Y}_1^{2^n}{\mathbb{F}_2}^{i^n(w)-1}: \Delta_{V^{s^n(w)}}(y^*) = 0 \right\rbrace \right) \xrightarrow[n\to \infty]{} 1,
\end{equation}
for any $w\in\mathcal{D}_0$, and 
\begin{equation}\label{eq:Delta_zero_submartingale_D1}
q_{W^{s^n(w)}}\left(\left\lbrace y^*\in\mathcal{Y}_1^{2^n}{\mathbb{F}_2}^{i^n(w)-1}: \Delta_{V^{s^n(w)}}(y^*) = 0 \right\rbrace \right) \xrightarrow[n\to \infty]{} 0,
\end{equation}
for any $w\in\mathcal{D}_1$.
\end{proposition}
\begin{IEEEproof}
By Lemma \ref{lem:Delta_zero_evolution} proved in the Appendix, the following inequality holds:
\begin{multline}
q_{W^-}\left(\left\lbrace y_1y_2\in\mathcal{Y}_1^{2}: \Delta_{V^-}(y_1y_2) = 0 \right\rbrace \right) + q_{W^+}\left(\left\lbrace y_1y_2u_1\in\mathcal{Y}_1^{2}\mathbb{F}_2: \Delta_{V^+}(y_1y_2u_1) = 0 \right\rbrace \right) \\
\geq 2 q_{W}\left(\left\lbrace y\in\mathcal{Y}: \Delta_{V}(y) = 0 \right\rbrace \right).
\end{multline}
By the recursive structure of the transformation, it follows that the process is a sub-martingale. The boundedness statement is trivial. Finally, one can verify that the limiting random variable is $\{0, 1\}$-valued in a similar fashion as in the  proof of Proposition \ref{prop:D_n_supermartingale} by using the expression in \eqref{eq:Delta_zero_minus} derived in Lemma \ref{lem:Delta_zero_evolution}:
\begin{multline}
\Expt \left[ \big\lvert q_{W^{s^{n+1}(w)}}\left(\left\lbrace y^*\in\mathcal{Y}_1^{2^n}{\mathbb{F}_2}^{i^n(w)-1}: \Delta_{V^{s^{n+1}(w)}}(y^*) = 0 \right\rbrace \right)\right. \\
- \left. q_{W^{s^n(w)}}\left(\left\lbrace y^*\in\mathcal{Y}_1^{2^n}{\mathbb{F}_2}^{i^n(w)-1}: \Delta_{V^{s^n(w)}}(y^*) = 0 \right\rbrace \right)\big\rvert \right]\\
\geq \frac{1}{2}  \Expt \left[ q_{W^{s^n(w)}}\left(\left\lbrace y^*\in\mathcal{Y}_1^{2^n}{\mathbb{F}_2}^{i^n(w)-1}: \Delta_{V^{s^n(w)}}(y^*) = 0 \right\rbrace \right) \right.\\ 
\times\left. \left( 1-  q_{W^{s^n(w)}}\left(\left\lbrace y^*\in\mathcal{Y}_1^{2^n}{\mathbb{F}_2}^{i^n(w)-1}: \Delta_{V^{s^n(w)}}(y^*) = 0 \right\rbrace \right) \right) \right]  \xrightarrow[n\to\infty]{} 0.
 \end{multline} 
 Finally, while \eqref{eq:Delta_zero_submartingale_D0} follows from \eqref{eq:T_n_0_case} upon noticing that $T(W)+q_{W}\left(\left\lbrace y\in\mathcal{Y}: \Delta_{V}(y) = 0 \right\rbrace \right) \leq 1$ holds,  \eqref{eq:Delta_zero_submartingale_D1} follows straightforwardly from \eqref{eq:D1-convergence-1}.
\end{IEEEproof}

Let us digress for a moment to look into the evolution of the parameter $I(W, V)$ under the polar transform. 
\begin{proposition}\label{prop:I_evolution}
The polar transform defined in \eqref{eq:def_pol_minus} and \eqref{eq:def_pol_plus}  preserves the mismatched mutual information of B-DMCs. i.e.,  
\begin{equation}\label{eq:I_evolution}
I(W^{-}, V^{-}) +  I(W^{+}, V^{+}) = 2I(W, V).
\end{equation}
\end{proposition}
\begin{IEEEproof}
In the Appendix, Lemmas \ref{lem:mismatch_I_minus} and \ref{lem:mismatch_I_plus} derive the expressions for $I(W^{-}, V^{-})$ and $I(W^{+}, V^{+})$, respectively. 
We can easily see that \eqref{eq:I_evolution} holds by using the expressions given in \eqref{eq:mismatch_I_minus} and \eqref{eq:mismatch_I_plus}, respectively, and the definition of $I(W, V)$ stated in \eqref{eq:mismatch_I_def_delta}.
\end{IEEEproof}
This result lays the foundation of another polar martingale, which we discuss next.
\begin{proposition}\label{prop:I_n_martingale}
The process $I_n(W, V)$ is a martingale, where $I_n(W, V)\leq 1$ holds for each $n\in\mathbb{N}$. Furthermore, the process converges a.s. toward a limiting random variable $I_\infty(W, V)$ such that $\Expt[I_\infty(W, V)] \geq I(W, V)$. 
\end{proposition}
\begin{IEEEproof}
The martingale property follows by Proposition \ref{prop:I_evolution} and the recursive structure of the transformation. The claim that $I_n(W, V)\leq 1$ holds, for each $n\in\mathbb{N}$, follows by the observation in  Remark \ref{rem:rem1-p5p6}. The claim about $I_\infty(W, V)$ holds by general results on bounded martingales upon noticing that $1-I_n(W, V)$ is a non-negative super-martingale, see for instance \cite[Cor. 11.7]{Martingales:Book}. 
\end{IEEEproof}

Now, we continue our investigation and refine the convergence result stated in \eqref{eq:D1-convergence-1}, for the paths $w\in \mathcal{D}_1$, with the help of the previous proposition.
\begin{proposition}\label{prop:Delta-convergence-1}
If $I(W, V) > -\infty$ holds, we have for all $\epsilon > 0$,
\begin{equation}\label{eq:D1-convergence-plusone}
W^{s^n(w)}\left(\left\lbrace y^*\in\mathcal{Y}_1^{2^n}{\mathbb{F}_2}^{i^n(w)-1}: 1-\Delta_{V^{s^n(w)}}(y^*) \geq \epsilon \right\rbrace \bigg\vert 0 \right) \xrightarrow[n\to \infty]{} 0,
\end{equation}
and
\begin{equation}\label{eq:D1-convergence-minusone}
W^{s^n(w)}\left(\left\lbrace y^*\in\mathcal{Y}_1^{2^n}{\mathbb{F}_2}^{i^n(w)-1}: 1+\Delta_{V^{s^n(w)}}(y^*) \geq \epsilon \right\rbrace \bigg\vert 1 \right) \xrightarrow[n\to \infty]{} 0,
\end{equation}
for any $w\in\mathcal{D}_1$. Thus, the process  $\Delta_{V^{s^n(w)}}\left(Y^*\right)$ measured under the distribution $W^{s^n(w)}(\cdot|0)$ converges in probability toward $1$, and the process  $\Delta_{V^{s^n(w)}}\left(Y^*\right)$ measured under the distribution $W^{s^n(w)}(\cdot|1)$ converges in probability toward $-1$.
\end{proposition}
\begin{IEEEproof} 
To analyze the convergence properties of the paths $w\in \mathcal{D}_1$ under the synthetic channel transition probabilities $W^{s^n(w)}(. |0)$, we proceed as follows: For each of the $N=2^n$ synthetic channels 
at the $n$-th stage of polarization, with $n\in\mathbb{N}$,  we compute the probability $W_N^{(i)}\left(\left\lbrace y_1^Nu_1^{(i-1)}: \Delta_{V_N^{(i)}}\left(y_1^Nu_1^{(i-1)}\right)\in\left[-1, -1+\xi\right) \right\rbrace \hspace*{1mm}\bigg|\hspace*{1mm}0\right)$, for any $\xi\in(0, 1)$,
and for $\beta\in(0,1)$, we let $M_n(\beta)$ be the fraction of channels for which this probability is larger than $\beta$, i.e., 
\begin{multline}
M_n(\beta) = \frac{1}{2^n}\#\Big\{i\in\{1, \ldots N = 2^n\} \ssep \\ W_N^{(i)}\left(\left\lbrace y_1^Nu_1^{(i-1)}: \Delta_{V_N^{(i)}}\left(y_1^Nu_1^{(i-1)}\right)\in\left[-1, -1+\xi\right) \right\rbrace \hspace*{1mm}\bigg|\hspace*{1mm}0\right) > \beta\Big\}.
\end{multline}
By Proposition \ref{prop:Delta-convergence-prob}, we know that
\begin{equation}
q_{W^{s^n(w)}}\left(\left\lbrace y^*\in\mathcal{Y}_1^{2^n}{\mathbb{F}_2}^{i^n(w)-1}: \lvert\Delta_{V^{s^n(w)}}(y^*)\rvert \in \left(-1+\xi,-1+\eta\right)  \right\rbrace \right) 
 \xrightarrow[n\to\infty] {} 0,
\end{equation}
for any $\xi < \eta \in(0, 1)$ and for any $w\in\Omega$ except possibly for a set of points of measure zero; thus, the limit of $M_n(\beta)$ is independent of the choice of $\xi$. Furthermore, by the martingale property of $I_n(W, V)$ discussed in Proposition \ref{prop:I_n_martingale}, we can write 
\begin{multline}
I_0 := I(W, V)
= \frac{1}{2N} \displaystyle\sum_{i=1}^{N} \sum_{\begin{subarray}{c}
			  y_{1}^{N}\in\mathcal{Y}^N \\
			  u_1^{(i-1)}\in\mathbb{F}_2
                    \end{subarray}} W_N^{(i)}\left(y_1^Nu_1^{(i-1)}\bigg|0\right)
\log\left(1+\Delta_{V_N^{(i)}}\left(y_1^Nu_1^{(i-1)}\right)\right) 
\\
+  \underbrace{\frac{1}{2N}\displaystyle\sum_{i=1}^{N}  \sum_{\begin{subarray}{c}
			  y_{1}^{N}\in\mathcal{Y}^N \\
			  u_1^{(i-1)}\in\mathbb{F}_2
                    \end{subarray}} W_N^{(i)}\left(y_1^Nu_1^{(i-1)}\bigg|1\right) \log\left(1-\Delta_{V_N^{(i)}}\left(y_1^Nu_1^{(i-1)}\right)\right)}_{\displaystyle\leq \frac{1}{2}\log 2}.
\end{multline}
Thus, 
\begin{multline}
 I_0 \leq \frac{1}{2}M_n(\beta) \left(\beta \log \xi + \log 2\right) 
+ \frac{1}{2}(1-M_n(\beta) )\log 2 + \frac{1}{2}\log 2 \\
\label{eq:I0_Mn_ineq}\leq \frac{1}{2}M_n(\beta) \beta \log \xi + \frac{3}{2} \log 2.
\end{multline}
By the remark that the limit of $M_n(\beta)$ is not changed by the choice of $\xi$, 
we conclude that for any $\xi>0$, $M_n(\beta)$ must vanish as $n\to\infty$, 
for otherwise, the right hand side of \eqref{eq:I0_Mn_ineq} will fail to be larger than $I_0> -\infty$ for small enough $\xi$. In view of the relation \eqref{eq:D1-convergence-1}, in Proposition  \ref{prop:Delta-convergence-prob}, we thus conclude  that, as long as $I_0> -\infty$, the only possibility when $w\in\mathcal{D}_1$ is to have  
\begin{equation}
W^{s^n(w)}\left(\left\lbrace y^*\in\mathcal{Y}_1^{2^n}{\mathbb{F}_2}^{i^n(w)-1}: \Delta_{V^{s^n(w)}}(y^*) \in\left(1-\xi, 1 \right]\right\rbrace \hspace*{1mm} \big|  \hspace*{1mm} 0\right)  \xrightarrow[n\to\infty] {} 1,
\end{equation}
for any $\xi\in(0, 1)$, This proves the claim in \eqref{eq:D1-convergence-plusone}. In a similar way, the argument can be repeated to prove the claim in \eqref{eq:D1-convergence-minusone} by showing that, when $w\in\mathcal{D}_1$, we must also have 
\begin{equation}
W^{s^n(w)}\left(\left\lbrace y^*\in\mathcal{Y}_1^{2^n}{\mathbb{F}_2}^{i^n(w)-1}: \Delta_{V^{s^n(w)}}(y^*) \in\left[-1, -1+\xi\right)\right\rbrace \hspace*{1mm} \big|  \hspace*{1mm} 1\right)  \xrightarrow[n\to\infty] {} 1,
\end{equation}
for any $\xi\in(0, 1)$, as long as $I_0> -\infty$. This completes the proof.
\end{IEEEproof}

Now, we are ready to prove the polarization theorem in the mismatched setting. 
\begin{IEEEproof}[Proof of Theorem \ref{thm:mismatched_polarization_D}]
Using the correspondence $L_V(y) = (1-\Delta_V(y))/(1+\Delta_V(y))$, the definition in \eqref{eq:mismatch_ML_Pe} can be equivalently written as
\begin{multline}
P_{\textnormal{e, ML}}(W, V) = \frac{1}{2} W\left(\{y\in\mathcal{Y}: \Delta_V(y)<0\}|0\right)+ \frac{1}{4}W\left(\{y\in\mathcal{Y}: \Delta_V(y)=0\}|0\right) \\
+ \frac{1}{2} W\left(\{y\in\mathcal{Y}: \Delta_V(y)>0\}|1\right) + \frac{1}{4}W\left(\{y\in\mathcal{Y}: \Delta_V(y)=0\}|1\right).
\end{multline}
Similarly, the genie-aided mismatched decoding error probability for the $i$-th synthetic channel can be written as
\begin{align}
&P_{\textnormal{e, ML}}(W_N^{(i)}, V_N^{(i)}) \nonumber\\
&= \frac{1}{2} W_N^{(i)}\left(\left\lbrace y_1^Nu_1^{(i-1)}\in\mathcal{Y}_1^{N}{\mathbb{F}_2}^{i-1}: \Delta_{V_N^{(i)}}\left(y_1^Nu_1^{(i-1)}\right) < 0 \right\rbrace \hspace*{1mm}\bigg|\hspace*{1mm}0\right) \nonumber\\
&+ \frac{1}{4} W_N^{(i)}\left(\left\lbrace y_1^Nu_1^{(i-1)}\in\mathcal{Y}_1^{N}{\mathbb{F}_2}^{i-1}: \Delta_{V_N^{(i)}}\left(y_1^Nu_1^{(i-1)}\right) = 0 \right\rbrace \hspace*{1mm}\bigg|\hspace*{1mm}0\right)  \nonumber\\
&+ \frac{1}{2} W_N^{(i)}\left(\left\lbrace y_1^Nu_1^{(i-1)}\in\mathcal{Y}_1^{N}{\mathbb{F}_2}^{i-1}: \Delta_{V_N^{(i)}}\left(y_1^Nu_1^{(i-1)}\right) > 0 \right\rbrace \hspace*{1mm}\bigg|\hspace*{1mm}1\right) \nonumber\\
&+ \frac{1}{4} W_N^{(i)}\left(\left\lbrace y_1^Nu_1^{(i-1)}\in\mathcal{Y}_1^{N}{\mathbb{F}_2}^{i-1}: \Delta_{V_N^{(i)}}\left(y_1^Nu_1^{(i-1)}\right) = 0 \right\rbrace \hspace*{1mm}\bigg|\hspace*{1mm}1\right)  .
\end{align}
It is now easy to see that the polarization results given in \eqref{eq:polarization_D_1}, \eqref{eq:polarization_D_2}, and \eqref{eq:polarization_D_3}, with the claim that $C_\mathrm{P}(W, V) = \Prob[D_\infty(W, V) = 1]$, are obtained as a corollary to Propositions \ref{prop:Delta_zero_submartingale} and \ref{prop:Delta-convergence-1}. In particular, by the relation \eqref{eq:Delta_zero_submartingale_D0} in Proposition \ref{prop:Delta_zero_submartingale},  it follows that $\lim_{n\to\infty}P_{\textnormal{e, ML}}(W^{s^n(w)}, V^{s^n(w)}) \to 1/2$ holds, for any $w\in\mathcal{D}_0$. On the other hand, since $I(W, V)$ is assumed to be finite in the hypothesis of the theorem, it follows by the relations \eqref{eq:D1-convergence-plusone} and \eqref{eq:D1-convergence-minusone} in Proposition \ref{prop:Delta-convergence-1} that $\lim_{n\to\infty} P_{\textnormal{e, ML}}(W^{s^n(w)}, V^{s^n(w)})$ is vanishing for any $w\in\mathcal{D}_1$. Consequently, the mismathed ML decoding error probability process $P_{\textnormal{e, ML}}(W^{s^n(w)}, V^{s^n(w)})$ converges a.s. to a limiting random variable taking values in $\{0, 1/2\}$.
\end{IEEEproof}

\subsection{Proof of Theorem  \ref{thm:rate_of_polarization_mismatched_setting}}\label{subsec:proof_rate_of_polarization_mismatched_setting}

We start by showing that the process $Z_n(W, V)$ satisfies the condition $(c.2)$ of Theorem \ref{thm:rate_of_polarization_Qn}. For that purpose, we look into the evolution of the    mismatched Bhattacharyya parameter under the one-step polar transform. 
\begin{proposition}\label{prop:Z_evolution}
For the polar transform defined in \eqref{eq:def_pol_minus} and \eqref{eq:def_pol_plus}, we have the following properties:
\begin{align}
\label{eq:Z_minus}&Z(W^{+}, V^{+}) = Z(W, V)^2, \\
\label{eq:Z_plus}&Z(W^{-}, V^{-}) \leq 2Z(W, V).
\end{align}
\end{proposition}
\begin{IEEEproof}
To prove the above relations, we write 
\begin{multline}
Z(W^+,V^+) =\displaystyle\sum_{y_1y_2u_1}\frac{1}{4}W(y_1|u_1)W(y_2|0)\sqrt{\displaystyle\frac{V(y_1|u_1\oplus 1)V(y_2|1)}{V(y_1|u_1)V(y_2|0)}} \\
+ \frac{1}{4}W(y_1|u_1\oplus 1)W(y_2|1)\sqrt{\displaystyle\frac{V(y_1|u_1)V(y_2|0)}{V(y_1|u_1\oplus 1)V(y_2|1)}} = Z(W, V)^2,
\end{multline}
and
\begin{multline}
Z(W^-,V^-) = \displaystyle\sum_{y_1y_2}\frac{1}{4}\left(W(y_1|0)W(y_2|0)+W(y_1|1)W(y_2|1)\right)\sqrt{\displaystyle\frac{L_V(y_1) + L_V(y_2)}{1+L_V(y_1) L_V(y_2)}} \\
+ \frac{1}{4}\left(W(y_1|1)W(y_2|0)+W(y_1|0)W(y_2|1)\right)\sqrt{\displaystyle\frac{1+L_V(y_1) L_V(y_2)}{L_V(y_1) + L_V(y_2)}} \leq 2Z(W, V),
\end{multline}
where we used the inequalities $\sqrt{(x+y)/(1+xy)} \leq \sqrt{x} + \sqrt{y}$, $\sqrt{(1+xy)/(x+y)} \leq \sqrt{1/x} + \sqrt{y}$, and their reciprocals by replacing $x \leftarrow 1/x$ and $y \leftarrow 1/y$, for $x, y\geq 0$.
\end{IEEEproof}

Next, we study the convergence properties of the process $Z_n(W, V)$ for the paths $w\in\mathcal{D}_1$ of the channel polarization process. 
\begin{corollary}\label{cor:Z_convergence_0}
Suppose $I(W, V) > -\infty$. In this case, $\lim_{n\to\infty}Z(W^{s^n(w)}, V^{s^n(w)}) = 0$, for all $w\in\mathcal{D}_1$. Consequently,  
 \begin{equation}\label{eq:Z_convergence_0}
 \Prob\left[Z_n(W, V)< \epsilon\right] \xrightarrow[n\to\infty] {} \Prob\left[D_\infty=1\right],
 \end{equation}
holds, for any $\epsilon \geq 0$.  
\end{corollary}
\begin{IEEEproof}
Suppose that $w\in\mathcal{D}_1$. We write
\begin{align}
&Z(W^{s^n(w)}, V^{s^n(w)}) \nonumber \\
&=  \frac{1}{2}\sum_{\begin{subarray}{c}
			  y^*\in\mathcal{Y}^{2^n}\times\mathbb{F}_2^{i^n(w)-1}: \\
			  \Delta_{V^{s^n(w)}}\left(y^*\right) > 1-\xi
                    \end{subarray}} W^{s^n(w)}\left(y^*|0\right)
\sqrt{L_{V^{s^n(w)}}(y^*)} \nonumber \\
&+ \frac{1}{2}\sum_{\begin{subarray}{c}
			  y^*\in\mathcal{Y}^{2^n}\times\mathbb{F}_2^{i^n(w)-1}: \\
			  \Delta_{V^{s^n(w)}}\left(y^*\right) \leq 1-\xi
                    \end{subarray}} W^{s^n(w)}\left(y^*|0\right)
\sqrt{L_{V^{s^n(w)}}(y^*)} \nonumber \\
&+ \frac{1}{2}\sum_{\begin{subarray}{c}
			  y^*\in\mathcal{Y}^{2^n}\times\mathbb{F}_2^{i^n(w)-1}: \\
			  \Delta_{V^{s^n(w)}}\left(y^*\right) < -1+\xi
                    \end{subarray}} W^{s^n(w)}\left(y^*|1\right)                 
\frac{1}{\sqrt{L_{V^{s^n(w)}}(y^*)}} \nonumber \\
\label{eq:sum-Z-2} 
&+ \frac{1}{2}\sum_{\begin{subarray}{c}
			  y^*\in\mathcal{Y}^{2^n}\times\mathbb{F}_2^{i^n(w)-1}: \\
			  \Delta_{V^{s^n(w)}}\left(y^*\right) \geq -1+\xi
                    \end{subarray}} W^{s^n(w)}\left(y^*|1\right)
\frac{1}{\sqrt{L_{V^{s^n(w)}}(y^*)}}
\end{align}
for any $\xi\in(0, 1)$.   By Remark \ref{rem:rem3-p3}, $Z(W^{s^n(w)}, V^{s^n(w)}) < \infty$ holds, for any $n\in\mathbb{N}$, since we have $I(W^{s^n(w)}, V^{s^n(w)}) > -\infty$. Moreover, using the relation
\begin{equation}
\sqrt{L_V(y)} = e^{-\arctanh{\Delta_V(y)}} =\sqrt{\frac{1-\Delta_V(y)}{1+\Delta_V(y)}},
\end{equation}
it is easy to see that if $\Delta_{V_N^{(i)}} \in  (1-\xi,1] $, we have
\begin{equation}
\sqrt{L_{V_N^{(i)}}} \in\left[0, \sqrt{\frac{\xi}{2-\xi}}\right),
\end{equation}
and if $\Delta_{V_N^{(i)}} \in [-1,-1+\xi)$, we have
\begin{equation}
\sqrt{\frac{1}{L_{V_N^{(i)}}}} \in\left[0, \sqrt{\frac{\xi}{2-\xi}}\right).
\end{equation} 
So, by Proposition \ref{prop:Delta-convergence-1}, we get 
\begin{align}
&\lim_{n\to\infty}Z(W^{s^n(w)}, V^{s^n(w)}) \nonumber\\
&\leq \sqrt{\frac{\xi}{2-\xi}} \nonumber\\
&+ \lim_{n\to\infty}\frac{1}{2}\sum_{\begin{subarray}{c}
			  y^*\in\mathcal{Y}^{2^n}\times\mathbb{F}_2^{i^n(w)-1}: \\
			  \Delta_{V^{s^n(w)}}\left(y^*\right) \leq 1-\xi
                    \end{subarray}} W^{s^n(w)}\left(y^*|0\right)
\sqrt{L_{V^{s^n(w)}}(y^*)}\nonumber\\
\label{eq:sum-Z-1}
&+ \lim_{n\to\infty}\frac{1}{2}\sum_{\begin{subarray}{c}
			  y^*\in\mathcal{Y}^{2^n}\times\mathbb{F}_2^{i^n(w)-1}: \\
			  \Delta_{V^{s^n(w)}}\left(y^*\right) \geq -1+\xi
                    \end{subarray}} W^{s^n(w)}\left(y^*|1\right)
\frac{1}{\sqrt{L_{V^{s^n(w)}}(y^*)}},
\end{align}
a.s., for any $\xi\in(0,1)$. Now, we claim that 
\begin{equation}\label{eq:sum-Z-convergent-1} 
\lim_{n\to\infty}\frac{1}{2}\sum_{\begin{subarray}{c}
			  y^*\in\mathcal{Y}^{2^n}\times\mathbb{F}_2^{i^n(w)-1}: \\
			  \Delta_{V^{s^n(w)}}\left(y^*\right) \leq 1-\xi
                    \end{subarray}} W^{s^n(w)}\left(y^*|0\right)
\sqrt{L_{V^{s^n(w)}}(y^*)} = 0
\end{equation}
and 
\begin{equation}\label{eq:sum-Z-convergent-2}
\lim_{n\to\infty}\frac{1}{2}\sum_{\begin{subarray}{c}
			  y^*\in\mathcal{Y}^{2^n}\times\mathbb{F}_2^{i^n(w)-1}: \\
			  \Delta_{V^{s^n(w)}}\left(y^*\right) \geq -1+\xi
                    \end{subarray}} W^{s^n(w)}\left(y^*|1\right)
\frac{1}{\sqrt{L_{V^{s^n(w)}}(y^*)}} = 0
\end{equation}
holds a.s,  for any $\xi\in(0,1)$. Note that $\sqrt{L_{V^{s^n(w)}}(y^*)}$ and $1/\sqrt{L_{V^{s^n(w)}}(y^*)}$ are continuous functions of $\Delta_{V^{s^n(w)}}(y^*)$. So, Proposition \ref{prop:Delta-convergence-1} and the continuous mapping theorem imply that the processes $\sqrt{L_{V^{s^n(w)}}(y^*)}$ and $1/\sqrt{L_{V^{s^n(w)}}(y^*)}$ converge in probability, both toward random variables taking the value $0$, the former under the distribution $W^{s^n(w)}\left(\cdot|0\right)$ and the latter under the distribution $W^{s^n(w)}\left(\cdot|1\right)$, for any $w\in\mathcal{D}_1$. Now, the limits in \eqref{eq:sum-Z-convergent-1} and \eqref{eq:sum-Z-convergent-2} must exist and be finite a.s., and in fact they must vanish a.s. as claimed, by the assumption that $I(W, V) > -\infty$ holds and as a corollary to the proof of Proposition \ref{prop:Delta-convergence-1}. As a result, we conclude that  $\lim_{n\to\infty}Z(W^{s^n(w)}, V^{s^n(w)}) \leq \sqrt{(\xi)/(2-\xi)}$ holds, for any $w\in\mathcal{D}_1$, and the claim of the proposition follows by taking $\xi\to 0^+$. 
\end{IEEEproof}


\begin{proposition}\label{prop:rate-of-convergence_0}
Let $Q_ n$ be a process such that:
\begin{itemize}
\item[$(rc.1)$] For each $n \in\mathbb{N}$, $Q_n\geq 0$, $Q_0$ is
constant, and $Q_n$ is measurable with respect to $\mathscr{F}_n$.
\item[$(rc.2)$] The condition $(c.2)$ of Theorem \ref{thm:rate_of_polarization_Qn} holds.
\item[$(rc.3)$] $\Prob\left[ Q_\infty(\omega) = 0 \vert \omega\in\mathcal{D}\right] = 1$, for some set $\mathcal{D}\subseteq \Omega$, and the probability $c:= \Prob[Q_\infty = 0]\in[0, 1]$ is well defined. In other words, the process $Q_n$ ``converges a.s. conditional on the event $\{ \omega\in\mathcal{D}\}$".
\end{itemize}
Then, for any $\beta\in(0, 1/2)$,
\begin{equation}\label{eq:Qn_convergence_1_relaxed}
\lim_{n\to\infty}\Prob[ Q_n < 2^{-2^{n\beta}}] = c.
\end{equation}
\end{proposition}
\begin{IEEEproof}
A careful reading of the proof of Theorem \ref{thm:rate_of_polarization_Qn} carried out in \cite[Sec. II]{5205856} reveals that relaxing \cite[conditions (z.1)--(z.3)]{5205856} to the above set of conditions does not change the validity of the proof. In other words, the claim in \eqref{eq:Qn_convergence_1_relaxed} can be proved following exactly the same proof steps presented in \cite[Sec. II]{5205856}. Note that the same conclusion can also be drawn by a careful reading of the proof presented in \cite[Proof of Lem. 2]{8052539}.
\end{IEEEproof}

Having all the necessary ingredients, we proceed with the proof of the theorem.
\begin{IEEEproof}[Proof of Theorem \ref{thm:rate_of_polarization_mismatched_setting}]
By Proposition \ref{prop:Z_evolution}, the condition $(c.2)$ of Theorem \ref{thm:rate_of_polarization_Qn} is satisfied for $Q_n:= Z_n(W, V)$ with $q=2$. Moreover,  by Remark \ref{rem:rem3-p3}, we know that $Z_n(W, V)\geq 0$, for any $n\in\mathbb{N}$, and we proved the convergence of the process to $0$ on the paths $w\in\mathcal{D}_1$ in Corollary \ref{cor:Z_convergence_0}. Therefore, the process  $Q_n:=Z_n(W, V)$ satisfies the conditions $(cr.1)$--$(cr.3)$ of Proposition \ref{prop:rate-of-convergence_0}  with $q=2$ and $c=\Prob\left[D_\infty=1\right]$, and we conclude that, for any $\beta\in(0, 1/2)$,
\begin{equation}\label{eq:fast_convergence_Z}
\lim_{n\to\infty}\Prob[ Z_n(W, V) < 2^{-2^{n\beta}}] = c,
\end{equation} 
where $c = \Prob[Z_\infty(W, V) = 0] = \Prob[D_\infty(W, V) = 1]$ holds. Finally, the proof of Theorem \ref{thm:rate_of_polarization_mismatched_setting} can be completed upon noticing that, by the observation given in Proposition \ref{prop:rem3-p1}, we have
\begin{equation}
\displaystyle\sum_{i\in\mathcal{A}_{N}} P_{\textnormal{e, ML}}(W_{N}^{(i)}, V_{N}^{(i)})  \leq \displaystyle\sum_{i\in\mathcal{A}_{N}} Z(W_{N}^{(i)}, V_{N}^{(i)}) = o(2^{-N^\beta}),
\end{equation}
for any $|\mathcal{A}_{N}| \geq NR$ such that $\mathcal{A}_{N}\subseteq\mathcal{A}_{N, \gamma}(W, V)$ given by \eqref{eq:info_set_mismatch} with the choice  $\gamma=o(2^{-N^\beta})$.
\end{IEEEproof}

\subsection{Proof of Theorem  \ref{thm:mismatch_family_of_LBs}}\label{subsec:proof_mismatch_family_of_LBs}
Although Proposition \ref{prop:I_n_martingale} tells us that $I_n(W, V)$ will converge a.s., it does not identify to which values the convergence will be. It is worth to include a remark on this issue. In the matched case $V=W$, we know that $I_{\infty}(W, W) = I_\infty(W) \in\{0, 1\}$ a.s. As a corollary to extremes of information combining~\cite{1412027}, it is known that among all channels $W$ with a given symmetric capacity $I(W)$, the BEC and the BSC polarize most and least in the sense of having the largest and the smallest differences between $I(W^+)$ and $I(W^-)$, see for instance \cite{6731577}. It follows by this extremality result that, for any $\gamma>0$,  there exists a $\xi>0$ such that
$|I(W) - I(W^-)| < \xi$ implies $I(W) \not\in (\gamma,1-\gamma)$. In the mismatched setting $V \neq W$, however,  it was observed in \cite[Sec. 7.2.3]{Mine/THESES} that it is possible to construct a counter-example to the analogous statement, namely that for any $\gamma>0$, there exists a $\xi>0$ such that
$|I(W,V) - I(W^-,V^-)| < \xi$ implies $I(W,V) \not\in (\gamma,1-\gamma)$. 

We now identify the convergence point of the process $I_n(W, V)$, for $n\in\mathbb{N}$, on the paths $w\in\mathcal{D}_1$. 
\begin{corollary}\label{cor:I-convergence-1}
If $I(W, V) > -\infty$, with probability one, the limiting random variable $I_\infty(W, V)=1$ whenever $D_\infty(W, V)=1$, i.e., $\lim_{n\to\infty}I(W^{s^n(w)}, V^{s^n(w)}) = 1$, for all $w\in\mathcal{D}_1$.
\end{corollary}
\begin{IEEEproof}
As in the proof of Corollary \ref{cor:Z_convergence_0}, we start by writing
\begin{align}
&I(W^{s^n(w)}, V^{s^n(w)}) \nonumber\\
&=  \frac{1}{2}\sum_{\begin{subarray}{c}
			  y^*\in\mathcal{Y}^{2^n}\times\mathbb{F}_2^{i^n(w)-1}: \\
			  \Delta_{V^{s^n(w)}}\left(y^*\right) > 1-\xi
                    \end{subarray}} W^{s^n(w)}\left(y^*|0\right)
\log\left(1+\Delta_{V^{s^n(w)}}\left(y^*\right)\right) 
\nonumber\\
&+ \frac{1}{2}\sum_{\begin{subarray}{c}
			  y^*\in\mathcal{Y}^{2^n}\times\mathbb{F}_2^{i^n(w)-1}: \\
			  \Delta_{V^{s^n(w)}}\left(y^*\right) \leq 1-\xi
                    \end{subarray}} W^{s^n(w)}\left(y^*|0\right)
\log\left(1+\Delta_{V^{s^n(w)}}\left(y^*\right)\right) 
\nonumber\\
&+ \frac{1}{2}\sum_{\begin{subarray}{c}
			  y^*\in\mathcal{Y}^{2^n}\times\mathbb{F}_2^{i^n(w)-1}: \\
			  \Delta_{V^{s^n(w)}}\left(y^*\right) < -1+\xi
                    \end{subarray}} W^{s^n(w)}\left(y^*|1\right) \log\left(1-\Delta_{V^{s^n(w)}}\left(y^*\right)\right)
\nonumber\\
\label{eq:sum-I-1}
&+ \frac{1}{2}\sum_{\begin{subarray}{c}
			  y^*\in\mathcal{Y}^{2^n}\times\mathbb{F}_2^{i^n(w)-1}: \\
			  \Delta_{V^{s^n(w)}}\left(y^*\right) \geq -1+\xi
                    \end{subarray}} W^{s^n(w)}\left(y^*|1\right) \log\left(1-\Delta_{V^{s^n(w)}}\left(y^*\right)\right)                    \end{align}
Thus,
\begin{align}
&I(W^{s^n(w)}, V^{s^n(w)}) \nonumber\\
\geq & \frac{1}{2}\log\left(2-\xi\right) \left(W^{s^n(w)}\left(y^*: \Delta_{V^{s^n(w)}}\left(y^*\right) > 1-\xi \big|0\right) + W^{s^n(w)}\left(y^*: \Delta_{V^{s^n(w)}}\left(y^*\right) < -1+\xi \big|1\right) \right)
\nonumber\\
&+ \frac{1}{2}\sum_{\begin{subarray}{c}
			  y^*\in\mathcal{Y}^{2^n}\times\mathbb{F}_2^{i^n(w)-1}: \\
			  \Delta_{V^{s^n(w)}}\left(y^*\right) \leq 1-\xi
                    \end{subarray}} W^{s^n(w)}\left(y^*|0\right)
\log\left(1+\Delta_{V^{s^n(w)}}\left(y^*\right)\right) 
\nonumber\\
\label{eq:sum-I-2}
&+ \frac{1}{2}\sum_{\begin{subarray}{c}
			  y^*\in\mathcal{Y}^{2^n}\times\mathbb{F}_2^{i^n(w)-1}: \\
			  \Delta_{V^{s^n(w)}}\left(y^*\right) \geq -1+\xi
                    \end{subarray}} W^{s^n(w)}\left(y^*|1\right) \log\left(1-\Delta_{V^{s^n(w)}}\left(y^*\right)\right),
\end{align}
for any $\xi\in(0, 1)$. Now we claim that
\begin{align}
\label{eq:sum-I-convergent-1} 
&\lim_{n\to\infty}\frac{1}{2}\sum_{\begin{subarray}{c}
			  y^*\in\mathcal{Y}^{2^n}\times\mathbb{F}_2^{i^n(w)-1}: \\
			  \Delta_{V^{s^n(w)}}\left(y^*\right) \leq 1-\xi
                    \end{subarray}} W^{s^n(w)}\left(y^*|0\right)
\log\left(1+\Delta_{V^{s^n(w)}}\left(y^*\right)\right)  = 0
\\
\label{eq:sum-I-convergent-2}
&\lim_{n\to\infty}\frac{1}{2}\sum_{\begin{subarray}{c}
			  y^*\in\mathcal{Y}^{2^n}\times\mathbb{F}_2^{i^n(w)-1}: \\
			  \Delta_{V^{s^n(w)}}\left(y^*\right) \geq -1+\xi
                    \end{subarray}} W^{s^n(w)}\left(y^*|1\right) \log\left(1-\Delta_{V^{s^n(w)}}\left(y^*\right)\right) = 0,
\end{align}
holds a.s., for any $w\in\mathcal{D}_1$. By Proposition \ref{prop:Delta-convergence-1}, we know that the process $\Delta_{V^{s^n(w)}}\left(Y^*\right)$ converges in probability toward a random variables taking the value $1$ under the distribution $W^{s^n(w)}\left(\cdot|0\right)$ and a random variables taking the value $-1$ under the distribution $W^{s^n(w)}\left(\cdot|1\right)$, for any $w\in\mathcal{D}_1$. Moreover, we observe that the limit of the sums in \eqref{eq:sum-I-convergent-1} and \eqref{eq:sum-I-convergent-2} must exist and they must be finite a.s., since from \eqref{eq:sum-I-1}, these sums are an integral part of the definition of the parameter $I(W^{s^n(w)}, V^{s^n(w)})> -\infty$ and we know by Proposition \ref{prop:I_n_martingale} that the limit $I_\infty(W, V) > -\infty$ exists a.s. We conclude that the limit of the sums in \eqref{eq:sum-I-convergent-1} and \eqref{eq:sum-I-convergent-2} equal zero for any $w\in\mathcal{D}_1$ by the assumption that $I(W, V) > -\infty$ holds and as a corollary to the proof of Proposition \ref{prop:Delta-convergence-1}. We thus get $\lim_{n\to\infty}I(W^{s^n(w)}, V^{s^n(w)}) \geq \log\left(2-\xi\right)$, for any $w\in\mathcal{D}_1$. The claim of the corollary follows by taking $\xi\to 0^+$. 
\end{IEEEproof}

Next, we prove the following technical result.
\begin{proposition}\label{prop:LB_C_P}
The limiting random variable $D_\infty(W, V)$ is such that:
\begin{equation}\label{eq:LB_C_P}
\Prob[D_\infty(W, V) = 1] \geq I(W, V).
\end{equation}
\end{proposition}
\begin{IEEEproof}
By Proposition \ref{prop:I_n_martingale}, we know that $\Expt[I_\infty(W, V)] \geq I(W, V)$ holds. By the relation given in Proposition \ref{prop:D_I_coupling}, we also know that $D_\infty(W, V) \geq I_\infty(W, V) \ln 2$ holds a.s. To see this, notice that $I_n(W, V) \leq \displaystyle\frac{1}{\ln 2}D_n(W, V)$ holds for each $n\in\mathbb{N}$, and so we can write
$\limsup_{n\to\infty}I_n(W, V) \leq \liminf_{n\to\infty}\displaystyle\frac{1}{\ln 2}D_n(W, V)$ a.s. to obtain the result. On the other hand, since by Proposition \ref{prop:D_n_supermartingale}, $D_\infty(W, V)$ is a.s. a $\{0,1\}$-valued random variable, we further notice that the bound $D_\infty(W, V) \geq I_\infty(W, V)$ must hold a.s.\footnote{We know that  $I_\infty(W, V)\leq 1$ always holds, and obviously, we have $I_\infty(W, V) \ln 2 \leq 0$ if and only if $I_\infty(W, V) \leq 0$.} Consequently,  $\Prob[D_\infty(W, V) = 1] = \Expt[D_\infty(W, V)] \geq \Expt[I_\infty(W, V)] \geq I(W, V)$, and we get the desired inequality.
\end{IEEEproof}

We are now ready to complete the proof of the theorem.
\begin{IEEEproof}[Proof of Theorem \ref{thm:mismatch_family_of_LBs}]
We have just proved in Proposition \ref{prop:LB_C_P} that 
\begin{equation}
\Prob[D_\infty(W, V)= 1] \geq I(W, V)
\end{equation} 
holds. Next, we discuss a trivial improvement to this lower bound. The above bound can be improved initially as $\Prob[D_\infty(W, V)= 1] \geq \big|I(W, V)\big|^+$, proving \eqref{eq:mismatch_family_of_LBs} for $n = 0$. Going one step further, we can improve the bound as follows:
\begin{equation}
\Prob[D_\infty(W, V)= 1] \geq \displaystyle\frac{1}{2} \big|I(W^-, V^-)\big|^+ + \displaystyle\frac{1}{2} \big|I(W^+, V^+)\big|^+,
\end{equation}
where we used the result of Proposition \ref{prop:I_evolution} which shows that the quantity $I(W, V)$ is preserved under the polar transform. More generally, the bound can be improved as
\begin{equation}
\Prob[D_\infty(W, V)= 1] \geq \displaystyle\frac{1}{2^n} \displaystyle\sum_{s^n\in\{+, -\}^n} \big|I(W^{s^n}, V^{s^n})\big|^+,
\end{equation}
for any $n\in\mathbb{N}$, by applying the same reasoning. This concludes the proof of the relation in \eqref{eq:mismatch_family_of_LBs}. The claim that the family of lower bounds in the right hand side of \eqref{eq:mismatch_family_of_LBs} converges to $C_P(W, V)$ is a direct consequence of Corollary \ref{cor:I-convergence-1}.
\end{IEEEproof}

\subsection{Proof of Theorem \ref{thm:rate_of_polarization_new}}\label{subsec:proof_rate_of_polarization_new}
For proving the final theorem, we first investigate the rate of polarization of the process $D_n(W, V)$ to zero.
\begin{proposition}\label{prop:rate_of_convergence_D_to_zero}
For any $\beta\in(0, 1/2)$, 
\begin{equation}\label{eq:rate_of_convergence_D_to_zero_1}
\lim_{n\to\infty}\Prob[ D_n(W, V) < 2^{-2^{n\beta}}] = 1-C_\mathrm{P}(W, V).
\end{equation}
\end{proposition}
\begin{IEEEproof}
The claim follows by \cite[Thms. 1 and 3]{5205856} which we restated in Theorem \ref{thm:rate_of_polarization_Qn}. To get \eqref{eq:rate_of_convergence_D_to_zero_1}, we simply observe that, by Remark \ref{rem:rem2-p4} and Propositions \ref{prop:D_evolution} and \ref{prop:D_n_supermartingale}, the conditions $(c.1)$, $(c.2)$, and $(c.3)$ of Theorem \ref{thm:rate_of_polarization_Qn} are satisfied for the process $Q_n:= D_n(W, V)$ with $q=2$ and $c = \Prob\left[D_\infty(W, V) = 0\right] = 1-C_\mathrm{P}(W, V)$.
\end{IEEEproof}
We can now proceed with the proof of the theorem.
\begin{IEEEproof}[Proof of Theorem \ref{thm:rate_of_polarization_new}]
In order to prove the theorem, we explore the relation between the limiting random variables $D_\infty(W)$ and $Z_\infty(W)$. It was proved in \cite[Prop. 9]{1669570} that $Z_\infty(W) \in\{0, 1\}$ a.s., and we proved in Proposition \ref{prop:D_n_supermartingale} that, similarly, $D_\infty(W) \in \{0, 1\}$ a.s. By  Proposition \ref{prop:D_Z_coupling}, the relation $D_\infty(W) + Z_\infty(W) \geq 1$ holds a.s., and thus, $D_\infty(W)=0$ if and only if $Z_\infty(W)=1$.  In view of these relations, the claim \eqref{eq:rate_of_polarization_new} of the theorem immediately follows from Proposition \ref{prop:rate_of_convergence_D_to_zero}. In order to prove the second claim, we consider the process $T_n(W)$. By Remark \ref{rem:rem2-p4}, Proposition \ref{prop:D_n_supermartingale}, and the relation given in \eqref{eq:ineq4}, it is clear that the process $Q_n:=T_n(W)$ satisfies the conditions $(c.1)$ and $(c.3)$ of Theorem \ref{thm:rate_of_polarization_Qn}, with $c = \Prob\left[T_\infty(W, V) = 0\right] = 1-C_\mathrm{P}(W, V)$ and $q=2$.  Furthermore, the condition $(\tilde{c}.2)$ of the theorem is also satisfied, since for each $n\in N$, 
\begin{align}
\label{eq:Tn_minus}&T_{n+1} = T_ n^2, &\quad \hbox{when } B_{n+1} = 0,\\
\label{eq:Tn_plus}&T_{n+1} \geq T_n, &\quad \hbox{when } B_{n+1} = 1,
\end{align}
where \eqref{eq:Tn_minus} follows from the product structure in \eqref{eq:d_minus}, and \eqref{eq:Tn_plus} from \cite[Lem. 6.16]{Mine/THESES}. Thus, by Theorem \ref{thm:rate_of_polarization_Qn}, if $T(W) >0$, then for any $\beta > 1/2$,
\begin{equation}\label{eq:rate_of_convergence_T_to_zero_2}
\lim_{n\to\infty}\Prob[ T_n(W) < 2^{-2^{n\beta}}] = 0.
\end{equation}
Upon noticing that $T_\infty(W)=0$ if and only if $Z_\infty(W)=1$, the claim in \eqref{eq:rate_of_polarization_new_2} follows. 
\end{IEEEproof}

\section{Final remarks}\label{sec:final_remarks}
Bridging the gap of theory with practice, by finding ``provably" capacity achieving codes that are also amenable to practice, has been a major practical concern of the field of coding theory \cite{4282117}. Ultimately, Ar{\i}kan \cite{1669570} came up  with the polar coding method which provided for the first time a code construction method demonstrated to achieve the channel capacity of B-DMCs with low complexity encoders and decoders \cite{7342879}. At the same time, considerations of decoding complexity also prompted the development of the mismatched decoding literature. However, until now, the study of the performance of sub-optimal decoders remained mainly focused on decoders with additive decision rules.  
In this paper, we introduced the mismatched polar decoder and studied its performance in a rigorous framework. We showed that the method of polar coding can be applied in mismatched communication scenarios. The study brought a new perspective to the literature of sub-optimal ``mismatched decoders" and led to the concept of the ``polar mismatched capacity". We are not aware of a similar investigation in the coding theory literature for other state-of-the-art coding schemes such as LDPC or Turbo codes. 

In particular, we presented a series of theorems which uncovered the following results about the polar mismatched capacity for any pairs of B-DMCs $W:\mathbb{F}_2\to\mathcal{Y}$ and $V:\mathbb{F}_2\to\mathcal{Y}$ such that $I(W, V)>-\infty$:
\begin{enumerate}
\item  The polar mismatch capacity is given by $C_\mathrm{P}(W, V)$ which is equal to the fraction of indices for which $D_n(W, V)$ converges to 1 as $n\to\infty$, 
\item $I(W, V)$ is a single letter lower bound on $C_\mathrm{P}(W, V)$,
\item This lower bound naturally generates a sequence of tighter lower bounds on $C_\mathrm{P}(W, V)$,
\item As in the matched case, the block decoding error probability of the mismatched polar decoder decays exponentially in the square root of the blocklength.
\end{enumerate}
We close the paper with a list of final remarks on the subject.

\subsection{Complexity}

Let us emphasize once more that polar coding with mismatched polar decoding uses the exact same encoding and decoding architectures proposed in \cite{1669570} for the original scheme. Therefore, as explained by Ar{\i}kan \cite[Thm. 5]{1669570}, these components of the system can be implemented in $O(N\log N)$ complexity as a function of the blocklength $N$. 

Regarding the code construction, we note that the original paper \cite[Sec. IX]{1669570} proposes a Monte Carlo based approach for the construction of the information sets of the form described in \eqref{eq:info_set_original}. Initially, this raised a complexity issue regarding the construction of polar codes. Though explicitly defined, tracking analytically the exact evolution of the synthetic channels in further steps is not possible in general since the output alphabets of the synthetic channels are growing exponentially fast and there is no general total ordering between the channels which hold for all B-DMCs. The problem of finding an efficient code construction algorithm for polar codes was first addressed by Mori and Tanaka in \cite{5205857} and \cite{5166430}, and later  these ideas were extended by Tal and Vardy \cite{6557004}  giving rise to an algorithm to carry out the computations approximately, but within guaranteed bounds, and efficiently. The particular case of Gaussian channels also received attention in a separate work \cite{6279525} which proposed using the Gaussian approximation for computing the bit error probabilities. Beside these studies, further considerations on  the complexity of polar codes revealed that the delay, construction, and decoding complexity can all be polynomially bounded as a function of the gap to capacity \cite{6960872}. 

As in the case of the original study  \cite{1669570}, we have not proposed an efficient algorithm to construct a polar code on the basis of the explicit definition of the information sets given in \eqref{eq:info_set_mismatch}.  Nevertheless, we believe it should also be possible in the mismatched setting to construct information sets of the form described in \eqref{eq:info_set_mismatch} via statistical methods. In addition, it might also be possible to adapt the low complexity code construction algorithm proposed in \cite{6557004} to the mismatched setting\footnote{Our claim is based on the observation that the algorithm developed in \cite{6557004} is based on the fusion of probability measures \cite[Chap. 1]{szekli1995stochastic} and the fact that the channels synthesized by the polar transform are ordered by the increasing convex ordering of their $\Delta$ random variables defined in \eqref{eq:delta}. See \cite[Def. 2]{6874810} for the definition of the increasing convex ordering, which is a partial order for B-DMCs ordering the information sets of polar codes \cite[Thm. 1]{6874810}, \cite[Cor. 6.17]{Mine/THESES}.}. This subject is left for future studies.

\subsection{Polarization as an Architecture to Boost the Classical Mismatched Capacity}
The motivation of the author to study the performance of `mismatched decoders' in the context of polar coding also led to the original study in \cite{6970855} which investigates whether channel polarization improves the classical mismatched capacity of B-DMCs. In the study \cite{6970855}, channel polarization has been proposed as a novel architecture to boost the classical mismatched capacity. 

We have already revisited in Section \ref{subsec:coding-theorem} an example given in that particular study. From Example \ref{ex:Example} and the works \cite{394641}, \cite{Balakirsky}, or \cite{Hui/THESES}, we readily see that the specific pairs of BSCs $W$ and $V$ with crossover probability $p\in[0, 0.5]$ and $(1-p)$, respectively, are such that $C(W, V) = 0$, while they satisfy  $C_\mathrm{P}(W, V)=I(W) > 0$. In fact, it is based on this example that \cite{6970855} draws the conclusion that there exist channels for which the sequence of polar transforms strictly improve the mismatched capacity of B-DMCs, and thus, it is possible to achieve communication rates higher than $C(W, V)$ by integrating the polarization architecture \cite{1669570} into the classical mismatched communication scenarios. However, \cite{6970855} also argues that the conclusion is not necessarily true in general, based on the results of a numerical experiment which reveals that the quantity $C(W, V)$ can be created or lost, and is not generally preserved, after applying the polar transform to the channels $W$ and $V$. Therefore, no general order between the polar mismatched capacity $C_\mathrm{P}(W, V)$ and the classical mismatched capacity $C(W, V)$ is formulated.  

A comparison between different sub-optimal decoders may not seem to  be necessarily fair, as each study corresponds to well-posed distinct mathematical problems. We stress, nevertheless, that the choice of the decoding scheme is part of the engineering problem. Since using ML decoding with the metric of a channel $V$ or using ML decoding with the metrics of $V^-$ first and then $V^+$ in a successive cancellation decoding configuration does not differ so much in complexity  for long sequences, we endorse the study \cite{6970855} which conveys the message that in some cases the mismatch at the decoder can be better exploited by using the polar transform architecture of Ar{\i}kan \cite{1669570}.
  
\subsection{List decoding of polar codes} 
Shortly after the inception of polar codes \cite{1669570}, it was discovered that the finite-length performance of the original scheme was not par with the state-of-the-art coding schemes implemented in wireless communications systems. This result motivated a great amount of research to focus on improving the finite-length performance of polar codes for their applicability in practical communication systems. List decoding was proposed in \cite{7055304} and shown to significantly improve the finite blocklength performance of polar codes. Upon this finding, the topic became one of the most active research areas in polar coding, and since, multiple list decoding algorithms and architectures have been proposed in the literature for the successive cancellation decoding of polar codes, see for instance \cite{7339660, 7339658, 7339671}. Let us, however, emphasize that the primary concern of these work is to show improvements experimentally, and in fact, results of theoretical nature on the performance of polar codes under list decoding are not currently available. 

Preliminary simulation results \cite{private:Mani} indicate that the finite-length performance of polar codes designed for mismatched communication scenarios might also improve when using a mismatched polar successive cancellation decoder implementation using a list decoding algorithm. In the view of the state-of-the-art practice, further investigations in this area will be important to  complement our work. The investigation could also help us understand the connection between the performance of polar coding with mismatched polar successive cancellation dcoding and mismatched ML decoding.

\section*{Acknowledgment}
The author would like to thank Emre Telatar for helpful discussions and is grateful to Vincent Y. F. Tan for his comments which helped to improve the presentation of this manuscript. This work was supported by Swiss National Science Foundation under grant number 200021-125347/1, and National University of Singapore grant number R-263-000-B37-133. 

\section*{Appendix}
In this appendix, we state and prove Lemmas \ref{lem:delta_plus}--\ref{lem:mismatch_I_plus}.

\begin{lemma}\label{lem:delta_plus}
Suppose $\Delta_1$, $\Delta_2$ are independent $[-1,1]$-valued
random variables with $\Expt\left[\sqrt{|\Delta_i|}\right]=\mu_i$.  Then
\begin{equation}\label{eq:dplus_1}
\Expt\Biggl[\sqrt{\biggl|\frac{\Delta_1+ \Delta_2}{1+\Delta_1\Delta_2}\biggr|}\Biggr]
\leq \mu_1+\mu_2-\mu_1\mu_2,
\end{equation}
and
\begin{equation}\label{eq:dplus_2}
\Expt\Biggl[\sqrt{\biggl|\frac{\Delta_1- \Delta_2}{1-\Delta_1\Delta_2}\biggr|}\Biggr]
\leq \mu_1+\mu_2-\mu_1\mu_2.
\end{equation}
\end{lemma}
\begin{IEEEproof}
If we take $a = \sqrt{\lvert\Delta_1\rvert}$ and $b = \sqrt{\lvert\Delta_2\rvert}$
in Lemma \ref{lem:ineq} which is given below, note that we get the following inequalities:
\begin{equation}\label{eq:dplus_ub1}
\sqrt{\biggl|\frac{\Delta_1+\Delta_2}{1+\Delta_1\Delta_2}\biggr|}
\leq \sqrt{|\Delta_1|}+\sqrt{|\Delta_2|}-\sqrt{|\Delta_1|}\sqrt{|\Delta_2|},
\end{equation}
and
\begin{equation}\label{eq:dplus_ub2}
\sqrt{\biggl|\frac{\Delta_1-\Delta_2}{1-\Delta_1\Delta_2}\biggr|}
\leq \sqrt{|\Delta_1|}+\sqrt{|\Delta_2|}-\sqrt{|\Delta_1|}\sqrt{|\Delta_2|}.
\end{equation}
Now, by taking expectations of both sides of \eqref{eq:dplus_ub1} and \eqref{eq:dplus_ub2}, and noting the independence of $\Delta_1$ and $\Delta_2$, the claims  \eqref{eq:dplus_1} and \eqref{eq:dplus_2} of the lemma follow.
\end{IEEEproof}

\begin{lemma}\label{lem:ineq}
For $a,b$ in the interval $[0,1]$,
\begin{equation}
\sqrt{\frac{|a^2-b^2|}{1-a^2b^2}} \leq \sqrt{\frac{a^2+b^2}{1+a^2b^2}} \leq a+b-ab.
\end{equation}
\end{lemma}
\begin{IEEEproof}
For the first inequality, we can assume without loss of generality that
$x=a^2\geq b^2=y$, and we only need to check
\begin{equation}
\frac{x-y}{1-xy}\leq \frac{x+y}{1+xy}, 
\end{equation}
for $x,y\in[0,1]$, or equivalently, $(x-y)(1+xy)\leq (x+y)(1-xy)$.  But this last
simplifies to $x^2y\leq y$, which clearly holds.

For the second inequality, squaring both sides and multiplying by
$(1+a^2b^2)$ we see that the inequality is equivalent to
\begin{equation}
(a+b-ab)^2(1+a^2b^2)-a^2-b^2 \geq 0.
\end{equation}
The left hand side factorizes as
$a(1-a)b(1-b)\bigl(2-ab(1+a+b-ab)\bigr)$.
Thus the lemma will be proved once we show that
\begin{equation}
t(1+s-t)\leq 2,
\end{equation}
where $s=a+b$ and $t=ab$.  Note that $0\leq s\leq 2$ and
$0\leq t \leq 1$.  Thus, $t(1+s-t)\leq t(3-t)\leq 2$.
\end{IEEEproof}

\begin{lemma}\label{lem:Delta_zero_evolution}
Let $W:\mathbb{F}_2\to\mathcal{Y}$ and $V:\mathbb{F}_2\to\mathcal{Y}$ be two B-DMCs. Then,
\begin{multline}\label{eq:Delta_zero_minus}
q_{W^-}\left(\left\lbrace y_1y_2\in\mathcal{Y}_1^{2}: \Delta_{V^-}(y_1y_2) = 0 \right\rbrace \right) \\
=  2q_{W}\left(\left\lbrace y\in\mathcal{Y}: \Delta_{V}(y) = 0 \right\rbrace \right) - q_{W}\left(\left\lbrace y\in\mathcal{Y}: \Delta_{V}(y) = 0 \right\rbrace \right)^2,
\end{multline}
and
\begin{equation}\label{eq:Delta_zero_plus}
q_{W^+}\left(\left\lbrace y_1y_2u_1\in\mathcal{Y}_1^{2}\mathbb{F}_2: \Delta_{V^+}(y_1y_2u_1) = 0 \right\rbrace \right)= q_{W}\left(\left\lbrace y\in\mathcal{Y}: \Delta_{V}(y) = 0 \right\rbrace \right)^2.
\end{equation}
\end{lemma}
\begin{IEEEproof}
Using the definition of the minus transformation in \eqref{eq:def_pol_minus} and the definition in \eqref{eq:d_minus}, we get
\begin{align}
& q_{W^-}\left(\left\lbrace y_1y_2\in\mathcal{Y}_1^{2}: \Delta_{V^-}(y_1y_2) = 0 \right\rbrace \right) \nonumber\\
&=\displaystyle\sum_{\begin{subarray}{c}
y_1y_2\colon \\ \Delta_{V^-}(y_1y_2) = 0 \end{subarray}}\displaystyle\frac{1}{2} W^{-}(y_1y_2|0)+\displaystyle\frac{1}{2} W^{-}(y_1y_2|1) \nonumber \\
& = \displaystyle\sum_{\begin{subarray}{c}
y_1y_2\colon \\ \Delta_{V^-}(y_1y_2) = 0 \end{subarray}} \displaystyle\frac{1}{4}\left(W(y_1|0)W(y_2|0)+W(y_1|1)W(y_2|1)\right)\nonumber \\
&\hspace{11mm} +\: \displaystyle\frac{1}{4}\left(W(y_1|1)W(y_2|0)+W(y_1|0)W(y_2|1)\right)\nonumber \\
& =  W\left(\left\lbrace y\in\mathcal{Y}: \Delta_{V}(y) = 0 \right\rbrace | 0\right)
+ W\left(\left\lbrace y\in\mathcal{Y}: \Delta_{V}(y) = 0 \right\rbrace | 1\right) \nonumber\\
&\hspace{11mm}- \displaystyle\frac{1}{4}\left(W\left(\left\lbrace y\in\mathcal{Y}: \Delta_{V}(y) = 0 \right\rbrace | 0\right)
+ W\left(\left\lbrace y\in\mathcal{Y}: \Delta_{V}(y) = 0 \right\rbrace | 1\right)\right)^2 \nonumber \\
& =  2q_{W}\left(\left\lbrace y\in\mathcal{Y}: \Delta_{V}(y) = 0 \right\rbrace \right) - q_{W}\left(\left\lbrace y\in\mathcal{Y}: \Delta_{V}(y) = 0 \right\rbrace \right)^2.
\end{align}
Using the definition of the plus transformation in \eqref{eq:def_pol_plus} and the definition in \eqref{eq:d_plus}, we get
\begin{align}
&q_{W^+}\left(\left\lbrace y_1y_2u_1\in\mathcal{Y}_1^{2}\mathbb{F}_2: \Delta_{V^+}(y_1y_2u_1) = 0 \right\rbrace \right) \nonumber\\
&= \displaystyle\sum_{\begin{subarray}{c}
y_1y_2u_1\colon \\ \Delta_{V^{+}}(y_1y_2u_1) = 0 \end{subarray}}\displaystyle\frac{1}{2} W^{+}(y_1y_2u_1|0)+\displaystyle\frac{1}{2} W^{+}(y_1y_2u_1|1)  \nonumber\\
&\geq \displaystyle\sum_{\begin{subarray}{c}
y_1\colon \\ \Delta_{V}(y_1) = 0 \end{subarray}}\displaystyle\sum_{\begin{subarray}{c}
y_2\colon \\ \Delta_{V}(y_2) = 0 \end{subarray}} \displaystyle\frac{1}{4}\left(W(y_1|0)W(y_2|0)+W(y_1|1)W(y_2|0)\right) \nonumber\\
&\hspace{10mm}+\: \displaystyle\sum_{\begin{subarray}{c}
y_1\colon \\ \Delta_{V}(y_1) = 0 \end{subarray}}\displaystyle\sum_{\begin{subarray}{c}
y_2\colon \\ \Delta_{V}(y_2) = 0 \end{subarray}} \displaystyle\frac{1}{4}\left(W(y_1|1)W(y_2|1)+W(y_1|0)W(y_2|1)\right) \nonumber\\
&\geq \displaystyle\frac{1}{4}\left(W\left(\left\lbrace y\in\mathcal{Y}: \Delta_{V}(y) = 0 \right\rbrace | 0\right)
+ W\left(\left\lbrace y\in\mathcal{Y}: \Delta_{V}(y) = 0 \right\rbrace | 1\right)\right)^2  \nonumber\\
&= q_{W}\left(\left\lbrace y\in\mathcal{Y}: \Delta_{V}(y) = 0 \right\rbrace \right)^2. 
\end{align}
\end{IEEEproof}

\begin{lemma}\label{lem:mismatch_I_minus}
Let $W:\mathbb{F}_2\to\mathcal{Y}$ and $V:\mathbb{F}_2\to\mathcal{Y}$ be two B-DMCs. Then,
\begin{align}
I(W^{-}, V^{-}) &= \frac{1}{4}\displaystyle\sum_{y_1y_2\in\mathcal{Y}^2} W(y_1|0)W(y_2|0)\log\left(1 + \Delta_{V}(y_1)\Delta_{V}(y_2)\right) \nonumber\\
& +\: \frac{1}{4}\displaystyle\sum_{y_1y_2\in\mathcal{Y}^2} W(y_1|0)W(y_2|1)\log\left(1 - \Delta_{V}(y_1)\Delta_{V}(y_2)\right) \nonumber\\
& +\: \frac{1}{4}\displaystyle\sum_{y_1y_2\in\mathcal{Y}^2} W(y_1|1)W(y_2|0)\log\left(1 - \Delta_{V}(y_1)\Delta_{V}(y_2)\right) \nonumber\\
\label{eq:mismatch_I_minus} & +\: \frac{1}{4}\displaystyle\sum_{y_1y_2\in\mathcal{Y}^2} W(y_1|1)W(y_2|1)\log\left(1 + \Delta_{V}(y_1)\Delta_{V}(y_2)\right). 
\end{align}
\end{lemma}
\begin{IEEEproof}
Using the definition of the minus transformation in \eqref{eq:def_pol_minus}, the proof follows upon observing that the following relations hold:
\begin{IEEEeqnarray}{rCl+x*}
1  + \Delta_{V}(y_1)\Delta_{V}(y_2) &=& \displaystyle\frac{2V(y_1|0)V(y_2|0) + 2V(y_1|1)V(y_2|1)}{\displaystyle\sum_{u\in\mathbb{F}_2}V(y_1|u)V(y_2|u)+V(y_1|u\oplus 1)V(y_2|u)}, \\
1 - \Delta_{V}(y_1)\Delta_{V}(y_2) &=& \displaystyle\frac{2V(y_1|0)V(y_2|1) + 2V(y_1|1)V(y_2|0)}{\displaystyle\sum_{u\in\mathbb{F}_2}V(y_1|u)V(y_2|u)+V(y_1|u\oplus 1)V(y_2|u)}. 
\end{IEEEeqnarray}
\end{IEEEproof}

\begin{lemma}\label{lem:mismatch_I_plus}
Let $W:\mathbb{F}_2\to\mathcal{Y}$ and $V:\mathbb{F}_2\to\mathcal{Y}$ be two B-DMCs. Then,
\begin{align}
I(W^{+}, V^{+}) &= \frac{1}{4}\displaystyle\sum_{y_1y_2\in\mathcal{Y}^2} W(y_1|0)W(y_2|0)\log\left(1 + \displaystyle\frac{\Delta_{V}(y_1) + \Delta_{V}(y_2)}{1 + \Delta_{V}(y_1)\Delta_{V}(y_2)}\right) \nonumber\\
& +\:\frac{1}{4}\displaystyle\sum_{y_1y_2\in\mathcal{Y}^2} W(y_1|0)W(y_2|1)\log\left(1 + \displaystyle\frac{\Delta_{V}(y_1) - \Delta_{V}(y_2)}{1 - \Delta_{V}(y_1)\Delta_{V}(y_2)}\right) \nonumber\\
& +\:\frac{1}{4}\displaystyle\sum_{y_1y_2\in\mathcal{Y}^2} W(y_1|1)W(y_2|0)\log\left(1 - \displaystyle\frac{\Delta_{V}(y_1) - \Delta_{V}(y_2)}{1 - \Delta_{V}(y_1)\Delta_{V}(y_2)}\right) \nonumber\\
\label{eq:mismatch_I_plus} & +\:\frac{1}{4}\displaystyle\sum_{y_1y_2\in\mathcal{Y}^2} W(y_1|1)W(y_2|1)\log\left(1 - \displaystyle\frac{\Delta_{V}(y_1) + \Delta_{V}(y_2)}{1 + \Delta_{V}(y_1)\Delta_{V}(y_2)}\right). 
\end{align}
\end{lemma}
\begin{IEEEproof}
Using the definition of the plus transformation in \eqref{eq:def_pol_plus}, the proof follows upon observing
\begin{IEEEeqnarray}{rCl+x*}
 1+\displaystyle\frac{\Delta_{V}(y_1)+\Delta_{V}(y_2)}{1+\Delta_{V}(y_1)\Delta_{V}(y_2)} &=& \displaystyle\frac{2V(y_1|0)V(y_2|0)}{\displaystyle\sum_{u\in\mathbb{F}_2}V(y_1|u)V(y_2|u)},\\
  1+\displaystyle\frac{\Delta_{V}(y_1)-\Delta_{V}(y_2)}{1-\Delta_{V}(y_1)\Delta_{V}(y_2)} &=& \displaystyle\frac{2V(y_1|0)V(y_2|1)}{\displaystyle\sum_{u\in\mathbb{F}_2}V(y_1|u)V(y_2|u\oplus 1)},\\
 1-\displaystyle\frac{\Delta_{V}(y_1)-\Delta_{V}(y_2)}{1-\Delta_{V}(y_1)\Delta_{V}(y_2)} &=& \displaystyle\frac{2V(y_1|1)V(y_2|0)}{\displaystyle\sum_{u\in\mathbb{F}_2}V(y_1|u)V(y_2|u\oplus 1)}. \\
  1-\displaystyle\frac{\Delta_{V}(y_1)+\Delta_{V}(y_2)}{1+\Delta_{V}(y_1)\Delta_{V}(y_2)} &=& \displaystyle\frac{2V(y_1|1)V(y_2|1)}{\displaystyle\sum_{u\in\mathbb{F}_2}V(y_1|u)V(y_2|u)}.
\end{IEEEeqnarray}
\end{IEEEproof}

\bibliographystyle{IEEEtran}


\end{document}